\RequirePackage{fix-cm}
\documentclass[smallcondensed]{svjour3}     % onecolumn (ditto)
\smartqed  % flush right qed marks, e.g. at end of proof
\usepackage{graphicx}
\usepackage[colorlinks=true,linkcolor=blue,citecolor=red,urlcolor=cyan]{hyperref}
\usepackage{amsmath}
\usepackage{amsfonts}
\usepackage{rotating} 
\usepackage{fix-cm}

\journalname{}

\begin{document}

%\title{Active Control of Boundary Layer Instabilities: Centralised vs. Decentralised approach%\thanks{Grants or other notes
\title{Centralised versus Decentralised Active Control of Boundary Layer Instabilities%\thanks{Grants or other notes
}
% \subtitle{Do you have a subtitle?\\ If so, write it here}

\titlerunning{Centralised vs. Decentralised control}        % if too long for running head

\author{R. Dadfar, N. Fabbiane, S. Bagheri and D. S. Henningson %etc.
}


\institute{S. Bagheri \at
              KTH Mechanics, Stockholm, SE-1044  \\
              Tel.: +46-8-7907161\\
              Fax: +46-8-7907161\\
              \email{shervin@mech.kth.se}           %  \\
           \and
%            S. Author \at
%               second address
}

\date{Received: date / Accepted: date}

\maketitle

\begin{abstract}

We use linear control theory to construct an output feedback controller for the  attenuation of small-amplitude Tollmien-Schlichting (TS) wavepackets in a flat-plate boundary layer. We distribute evenly in the spanwise direction up to $72$ localized objects near the wall ($18$ disturbances sources, $18$ �actuators, $18$ estimation sensors and $18$ objective sensors).
In a fully three-dimensional configuration, the interconnection between inputs and outputs becomes quickly unfeasible when the number of actuators and sensors increases in the spanswise direction. The objective of this work is to understand how an efficient controller may be designed by connecting only a subset of the actuators to sensors, thereby reducing the complexity of the controller, without comprising the efficiency. We find that using a semi-decentralized approach -- where small {\it control units} consisting of 3 estimation sensors connected to $3$ actuators are replicated $6$ times along the spanwise direction --  results only in a $11\%$ reduction of control performance. %The key parameter that the control unit size depends on, is the angle of which the TS wavepackets spread in the spanwise direction.
Our results reveal that the best performance is obtained for a control unit which (i) is sufficiently ``wide'' to account for the full spanwise scale of the wavepacket when it reaches the actuators and (ii) is designed to account for the perturbations which are coming from the lateral sides (crosstalk) of the estimation sensors. We have also found that the influence of  crosstalk is not as essential as the spreading effect.

\end{abstract}
%%%%%%%%%%%%%%%%%%%%%%%%%%%
%                              Introduction
%%%%%%%%%%%%%%%%%%%%%%%%%%%
\section{Introduction}
% \cite{bagheri2011transition}
Drag reduction methodologies in vehicles and aircrafts have received considerable attention during the past decades \cite{thomas1984aircraft}. These techniques provide the possibility to significantly reduce the operational cost in transportation sector and also improve the environmental consequences.  In boundary layer flows, drag reduction can be achieved by extending the laminar region on the aerodynamics parts of vehicles by delaying the transition from laminar to turbulence. Among different techniques to delay the transition, active control is of a great importance. This approach adds external energy to the system in terms of predetermined actuation (open loop) or on-line computation of the actuation law using feedback information from the measurement sensors (reactive control). One particular reactive control strategy employed in this study is output feedback control\cite{doyle89}, where the actuation is determined by measuring external disturbances.

In an environment characterised by low turbulence levels, two-dimensional perturbations -- Tollmien-Schlichting (TS) -- wavepackets are triggered inside the boundary layer. The TS waves grow exponentially in amplitude as they move downstream until a point where nonlinear effects are significant and transition to turbulence is triggered.  An important trait of this transition scenario, which also enables the use of linear control theory, is that the initial stage of the perturbation growth inside the boundary layer is well described by a linear system. Moreover, due to the large sensitivity of such flows to an external excitation, one can  influence the TS waves by introducing small local perturbation in small region of the flow via proper localised devices requiring minute energy. 
There is now substantial literature where linear control theory is combined with numerical simulations to control transition in wall-bounded flows. Pioneering work include the control of Orr-Sommerfeld equations \cite{joshi1997systems}, distributed control using convolution kernels \cite{cortelezzi1998robust,hogberg:bewley:henning:03:b} and a localized control approach \cite{bagheri:brandt:henning:09}. The term {\it localized} �in the latter approach refers to the use of a limited number of small  compact actuation and estimation devices positioned in specific manner  to allow efficient control. The fact that the number of inputs/outputs ($\mathcal{O}(10)$) is order of magnitudes smaller than the dimensions of flow system ($\mathcal{O}(10^7)$) provides amenable conditions for reducing the order of the system by constructing  a low-dimensional model (ROM). Here, we report on our most advanced configuration (placing up to 72 inputs/outputs) so far. The configuration of our numerical system is chosen as to resemble as closely as possible the experimental study performed by \cite{2006:li:gaster}.  This investigation extends or complements our previous work on two-dimensional disturbances using blowing/suction and shear stress measurements \cite{bagheri_aiaa}, three-dimensional linear \cite{2010:semeraro:bagheri:brandt:henningson} and nonlinear \cite{2013:semeraro:bagheri:brandt:henningson} investigations. Relevant reviews on this subject  are provided in \cite{Bewley01,kim:bewley:07,bagheri2011transition}.  

We will report on the efficiency of a centralised and  a decentralised control strategy \cite{glad:ljung,lewis}. In the former approach all the sensors are connected to all the actuators. Since the complexity of a  controller is related to the number of interconnections, this approach becomes unfeasible when reaching  $\mathcal{O}(10^2)$ inputs and outputs. This is certainly a restrictive issue, since in a localized control approach the number of required sensors and actuators increase with the span of the plate. A solution to this restriction is a decentralised controller where one disregards some of the interconnections which are not essential to the dynamics of the system. Then one replicates the same controller (called control unit) along the span of the system to cover a larger spanwise distance. In this study, several different control units are designed and their performances are compared.

%%%%%%%%%%%%%%%%%%%%%%%%%%%
%                             Configuration
%%%%%%%%%%%%%%%%%%%%%%%%%%%
\section{Flow and Control Configuration}
\subsection{Governing equations}
The dynamics and control of small-amplitude perturbations in a viscous, incompressible flow developing over an unswept flat plate are investigated using direct numerical simulation (DNS). The disturbance dynamics is governed by the Navier-Stokes equation linearised around a spatially developing zero-pressure-gradient boundary layer flow as

\begin{subequations}\label{equ:LS}
\begin{alignat}{2}
 &\frac{\partial{{u}}}{\partial{t}} = -({U}\cdot \nabla ){u} -({u} \cdot \nabla){U}
 -\nabla{p}+\frac{1}{Re}\nabla ^2 {u}+\lambda_f (x)u,   \\
& \nabla \cdot {u} =0,               \\
& {u} ={u}_0 \quad \text{at} \quad t=t_0,
\end{alignat}
\end{subequations}
where the disturbance velocity and pressure fields are denoted by ${u}(x,y,z,t)$ and ${p}(x,y,z,t)$; $x$, $y$ and $z$  denote the streamwise, wall normal and spanwise direction, respectively. Furthermore, ${U(x,y)}$ and  ${P(x,y)}$ represent the baseflow velocity and pressure; they are a solution to the steady, nonlinear Navier-Stokes equation. In this study, all the spatial coordinates are normalised with the displacement thickness $\delta^*$ at the inlet of the computational box. The Reynolds number is defined  based on the displacement thickness as $Re = U_{\infty}\delta^*/\nu$ where the $U_\infty$ denotes the uniform free stream velocity and $\nu$ is the kinematic viscosity; all the simulations are performed at $Re=915$ which correspond to a distance of $312 \delta^*$ from the origin of the plate to the inlet of the computational box. The no-slip boundary condition is considered at the wall ($y=0)$, while Dirichlet boundary condition with vanishing velocity is employed at the upper boundary $(y=L_y)$; this boundary condition is applied far enough from the boundary layer to ensure negligible influence on the dynamics of the perturbations. Periodicity is assumed in the spanwise and streamwise directions. In the latter, the term $\lambda(x)$ is implemented to enforce this periodicity so that a spectral Fourier expansion techniques can be employed. The function $\lambda(x)$ is zero inside the physically relevant part of the domain where the dynamics are investigated and has nonzero value at the end of the domain where a fringe region is applied \cite{nordstrom1999fringe}. The simulation is performed using a pseudo-spectral DNS code \cite{chevalier2007simson} where Fourier series are employed in the wall-parallel directions and the wall-normal direction is expanded in Chebyshev polynomials.
%--------------------------------------- config ------------
\begin{figure}%[t!]
\begin{center}
\includegraphics[width=0.8\textwidth]{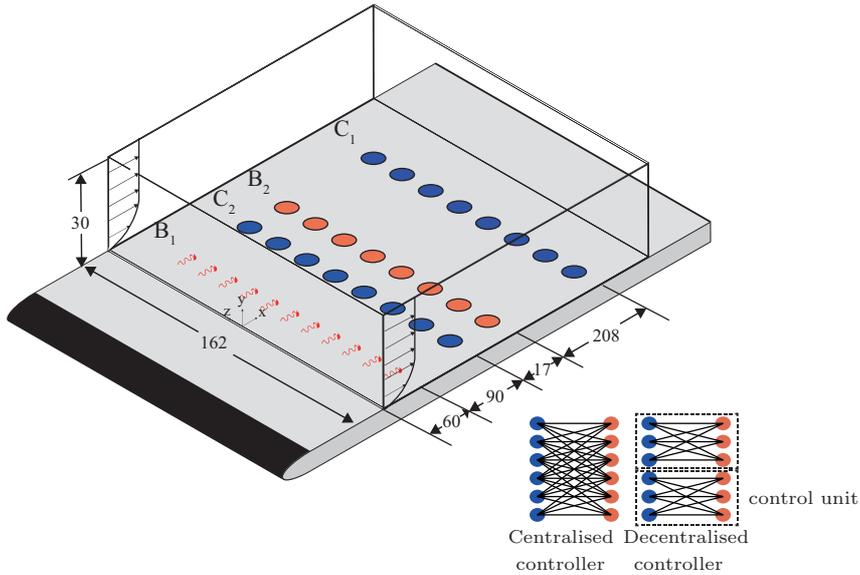} 
\put(-170,120){}
\put(-210,80){}
\put(-90,10){\scriptsize Centralised}
\put(-87,0){\scriptsize controller}
\put(-47,10){\scriptsize Decentralised}
\put(-43,0){\scriptsize controller}
\put(0,25){\scriptsize control unit}
\end{center}
\caption{Input-output configuration of the system. The input $B_1$ is a row of localised disturbances located at $x = 60$, convected downstream and converted to a TS wavepacket. The control action is provided by the input $B_2$, consists of a row of actuators located at $x = 167$. A set of localised estimation sensors, at $x = 150$ upstream of the actuator is employed. A row of output sensors at $x=375$ is implemented as the objective function of the controller. Two control strategies, centralised and decentralised are used. In the former all the sensors and actuators are wired together while in the latter, a control unit with a limited interconnections is designed and replicated along the span. There are in total $18$ disturbances $B_1$, $18$ sensors $C_2$, $18$ actuators $B_2$ and $18$ outputs $C_1$. Only $8$ of those are depicted in the figure.}
\label{fig:config}
\end{figure}

%%%%%%%%%%%%%%%%%%%%%%%%%%%%%%%%%%%%%%%%%%%%%%%%%%%%%%%%%%%%%%%%%%%%%%%%%%%%%%%%
%                             Input-Output System
%%%%%%%%%%%%%%%%%%%%%%%%%%%%%%%%%%%%%%%%%%%%%%%%%%%%%%%%%%%%%%%%%%%%%%%%%%%%%%%%
\subsection{Input-Output System}
A schematic representation of the input-output configuration is depicted in Fig.~\ref{fig:config}. The linearised Navier-Stokes equation with inputs and outputs can be written is state space form as  
\begin{subequations}\label{eq:sys}
\begin{alignat}{2}
 \dot{{u}}(t) &= {A}{u}(t)+{B_1} w(t)+{B_2} \phi(t),\\  
{v}(t)&={C_2}{u}(t)+ \alpha g(t),\\  
{z}(t)&=  \binom{C_1}{0}  u(t) + \binom{0}{R^{1/2}} \phi(t). 
\end{alignat}
\end{subequations}
Henceforth, $u(t)\in\mathbb{R}^n$  denotes the state vector, whereas $w(t)\in\mathbb{R}^{d}, \phi(t) \in \mathbb{R}^m, v(t)\in\mathbb{R}^p, g(t)\in \mathbb{R}$ and $z(t) \in \mathbb{R}^k$ denote time signals.
 The matrix $A \in \mathbb{R}^{n \times n}$ represents the linearised and spatially discretised Navier-Stokes equation. The above form has been reported in numerous works (see e.g. \cite{iutam09onofrio}) and only a short description is provided here:
\begin{itemize}
\item The first input ($B_1 w(t)$) is composed of ${B_1}\in\mathbb{R}^{n\times d}$ representing the spatial distribution of $d$ localised disturbances located at the upstream end of the domain and  white noise signals $w(t) \in\mathbb{R}^{d}$. These inputs represent a model of perturbations introduced inside the boundary layer by \emph{e.g} roughness and free-stream perturbations. 
\item In the second input ($B_2\phi(t)$),  ${B_2}\in\mathbb{R}^{n\times m}$ represents the spatial support of $m$ actuators located inside the boundary layer near the wall. They are fed by the control signal $\phi(t) \in \mathbb{R}^m$, which is to be determined by an appropriate controller. 
\item The $p$ output measurement provided by $v(t)\in\mathbb{R}^p$ detect information about the travelling structures  by the localised sensors ${C}_2\in \mathbb{R}^{p\times n}$. These measurements are corrupted by ${\alpha} g(t)$. More precisely, $g(t)\in \mathbb{R}^p$ is a white noise signal and ${\alpha}$ the level of noise.
% 
%  and $I_{\alpha} g(t)$, with $\alpha$ on the diagonal entries is a model of different noise level corrupting the sensors measurements; $I_{\alpha} \in \mathbb{R}^{p \times m}$.
\item  The output $z(t) \in \mathbb{R}^k$  extracts information from the flow in order to evaluate the performance of the controller. This is done by localised outputs $C_1 \in \mathbb{R}^{k \times n}$ with a spatial distribution located far downstream in the computational box. 
%It is used to design the controller and can be applied to assess the performance of the controller. 
In fact, the minimisation of the output signal detected by $C_1$ is the objective of our LQG controller; the aim is to find a control signal $\phi(t)$ able to attenuate the amplitude of the disturbance detected by $C_1$. Hence, the objective function reads
\begin{align}\label{eq:cost}
\mathcal{E}\left ({\| z \|^2_{L^2_{[0,\infty]}}}\right ) =\mathcal{E}\left \{ {\int_0^\infty {u}^T{C}^T_{1}{C}_{1}{u} +\phi^T R\phi \hspace{0.2cm} \mathrm{d}t } \right\},
\end{align}
%
% \begin{equation}\lablel{eq:objective}
% \| z\| ^2_{L^2_{[0 , \infty ]} } = \int_0^{\infty} \! u^T {C}_1^TC_1{u} +  \phi^TI_lI_l \phi \, \mathrm{d} t.
% \label{eq:system_norm}
% \end{equation}
where $\mathcal{E}(\cdot)$ is the expectation operator.
The matrix $R \in \mathbb{R}^{k \times m}$ contains the control penalty $l^2$ in each diagonal entry and represents the expense of the control. This parameter is introduced as a regularisation term accounting for physical restrictions. 
Large values of control penalty results in weak actuation and creates low amplitude control signal whereas low values of control penalty leads to strong actuation. 

\end{itemize}

Following \cite{2010:semeraro:bagheri:brandt:henningson}, we define the spatial distribution of the sensors and actuators with  a Gaussian divergence-free function as
\begin{align}
h(x,y,z)=a
\left(
\begin{array}{c}
 \sigma_x \gamma_y\\
-\sigma_y \gamma_x \\
 0
\end{array}
\right) e^{-\gamma_x^2-\gamma_y^2-\gamma_z^2},
\label{equ:exp}
\end{align}
where
\begin{align}\label{eq:eq}
\gamma_x=\frac{x-x_0}{\sigma_x}, \quad \gamma_y=\frac{y-y_0}{\sigma_y}, \quad \gamma_z=\frac{z-z_0}{\sigma_z},
\end{align}
and  $(x_0, y_0,z_0)$ is the centre of the Gaussian distribution. The scalar quantities $(\sigma_x, \sigma_y, \sigma_z)$ represent the corresponding size (values given  in Tab. \ref{tab:elements}). The scalar $a$ represents an amplitude which is equal to $2 \times 10^{-3}$ for the actuators and one for the sensors.
%
%------------------------------------------ specification--------------------
\begin{table}
\begin{tabular}{c c c c c}
& & & &  \\ % put some space after the caption
\hline
\hline
Element  & Symbol & Number & Location & Parameters \\
 $ - $& $ - $       & & $(x_0,y_0)$ & $(\sigma_x,\sigma_y,\sigma_z)$\\

\hline
Disturbances& $B_1$ &$ 18 $ &$ (60,0)$ &$(6,1.5,8)$ \\
Sensors     & $C_2$ &$ 18 $ &$(150,0)$ &$(2,1.5,2)$\\
Actuators   & $B_2$ &$ 18 $ &$(167,0)$ &$(6,1.5,8)$\\
Outputs     & $C_1$ &$ 18 $ &$(375,0)$ &$(5,1.5,6)$\\
\hline
\hline
\end{tabular}
\centering
\caption{The main parameters characterising the spatial distribution of the sensors and the actuators. All the elements are located at $z_0=-76.5$ and distributed along the span with the spanwise spacing $\Delta_z=9$.} 
\label{tab:elements}
\end{table}
Most of our simulation is performed for the setup reported in Tab. \ref{tab:elements}.  We denote the $ith$ element of the disturbance vector $ B_{1}$ by $B_{1,i}$ corresponding to the signal $w_i(t)$.

%%%%%%%%%%%%%%%%%%%%%%%%%%%%%%%%%%%%%%%%%%%%%%%%%%%%%%%%%%%%%%%%%%%%%%%%%%%%%%%%%%%%%
%                            Model Reduction and Control Design
%%%%%%%%%%%%%%%%%%%%%%%%%%%%%%%%%%%%%%%%%%%%%%%%%%%%%%%%%%%%%%%%%%%%%%%%%%%%%%%%%%%%%
%
% Finally, for convenience, we introduce the compact form of the state-space system,
% %
% \begin{subequations}\label{equ:sys}
% \begin{alignat}{2}
% \dot{u} &= {A} {u} + {B} {f},  \\
% y&={C} {u} + D f,
% \end{alignat}
% \end{subequations}
% where  
% \begin{equation*}\label{equ:sys4}
% B= (B_1,0,B_2), \hspace{0.5cm}
% C = \begin{pmatrix}
%    C_1  \\
%    C_2
% \end{pmatrix},
% \hspace{0.5cm}  
% D=
% \begin{pmatrix}
%    0 & 0 & I_l  \\
%    0 & I_\alpha & 0
% \end{pmatrix}, 
% \end{equation*}
% and 
% \begin{equation*}\label{equ:sys5}
% f(t) = \begin{pmatrix}
%    w(t)  \\
%    g(t)  \\
%   \phi(t)
% \end{pmatrix}, 
% \hspace{0.5cm} 
% y(t) = \begin{pmatrix}
%    z(t)  \\
%    v(t)  
% \end{pmatrix}. \nonumber
% \end{equation*}

% This system has a large number of degree of freedom (DOF) - order of $10^7$. Hence, it is difficult to apply the standard methodology for the control design in an efficient manner. This restriction can be addressed by designing a low-dimensional system that preserves the essential dynamics of the original system.
\begin{figure}%[t!]
\begin{center}
\includegraphics[width=0.77\textwidth]{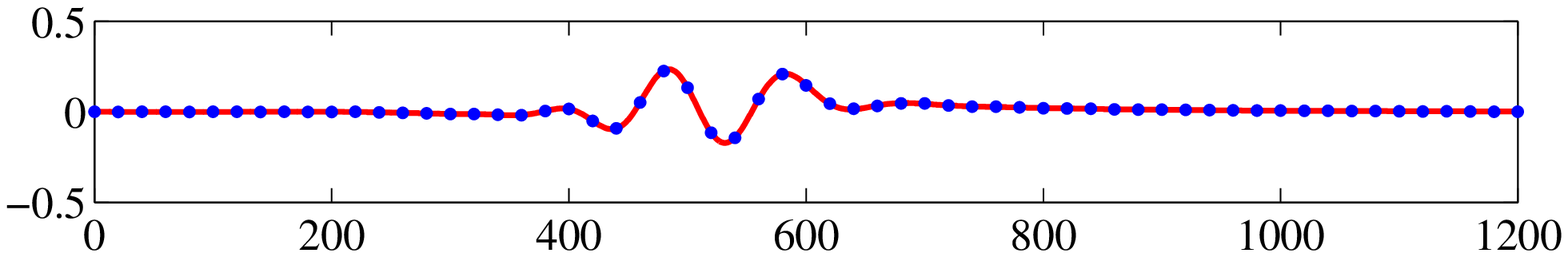} 
% \put(-305,0){{\begin{sideways}{${B}_{2,8} w \rightarrow {C}_{1,8}{u}$}  \end{sideways}}} 
\put(-280,40){(a)} \\
\includegraphics[width=0.77\textwidth]{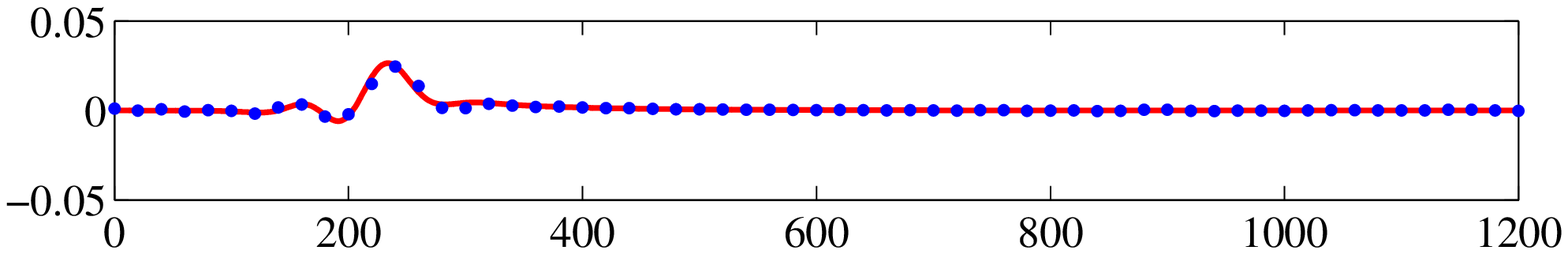} 
% \put(-305,0){{\begin{sideways}{${B}_{1,8} w \rightarrow {C}_{2,8}{u}$}  \end{sideways}}} 
\put(-280,40){(b)} \\
\includegraphics[width=0.77\textwidth]{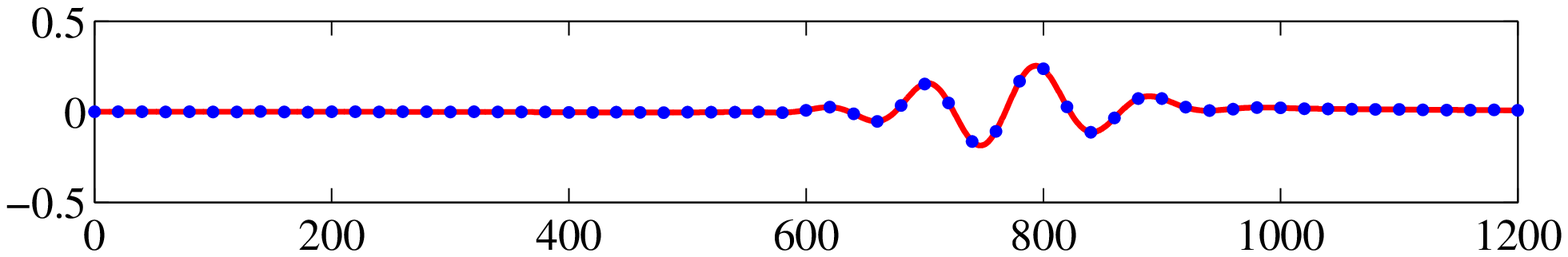} 
% \put(-305,0){{\begin{sideways}{${B}_{1,8} w \rightarrow {C}_{1,8}{u}$}  \end{sideways}}} 
\put(-280,40){(c)} 
\put(-135,-10){$t$}
\end{center}
\caption{Impulse response of the system (a) from the input $B_{2,8}$ to the output $C_{1,8}$, (b) from the input $B_{1,8}$ to the output $C_{2,8}$  and (c) from the input $B_{1,8}$ to the output $C_{1,8}$. The red line shows the DNS results, while the dotted lines indicates the impulse response of the reduced model (Case N Tab. \ref{tab:pf}) }
\label{fig:mor}
\end{figure}

\subsection{Model Reduction} 
We construct a reduced-order model of the system by projecting the $n-$dimensional state onto a low-dimensional subspace of dimension $r$. Expanding the state in a linear combination of columns of the expansion basis $\Phi=(\phi_1,\phi_2,\cdots \phi_r) \in \mathbb{R}^{n\times r}$ as
\begin{subequations}\label{equ:mor}
\begin{alignat}{2}
&  u = \Phi \hat{u}   \\
&  \hat{u} = \Psi^T u,
\end{alignat}
\end{subequations}
where $\Psi=(\psi_1,\psi_2,\cdots \psi_r) \in \mathbb{R}^{n\times r}$ are the adjoint modes, bi-orthogonal to the expansion basis $\Phi$, i.e. $\Psi^T \Phi = I$. Substituting Eq. (\ref{equ:mor}a) into the system Eq. (\ref{eq:sys}) and using the bi-orthogonality of the basis, the reduced system of order $r$ is obtained as
%----------------------------------------------  reduced system
\begin{subequations}\label{equ:mor2}
\begin{alignat}{2}
 &A_r =  \Psi^TA\Phi, \\
 &B_{1r} =  \Psi^TB_1,  \hspace{0.7cm} B_{2r} =  \Psi^TB_2, \\
 &C_{1r} =  C_1 \Phi,   \hspace{1cm} C_{2r} =  C_2 \Phi. 
\end{alignat}
\end{subequations}
%----------------------------------------------  end reduced system
The choice of the basis function is crucial for the performance of the reduced order system \cite{bagheri:hoepffner:schmid:henning:09,2009:barbagallo:sipp:schmid}.
 We use a balanced-mode-basis \cite{moore1981principal,willcox:peraire:02,rowley2005model} that preserves the dynamics between the inputs and outputs of the system. The states that are equally observable and controllable form a hierarchy of so-called balanced modes. The method is based on the concepts of observability and controllability\cite{zhou:doyle:glover:02}, which provide a means to characterize the states in terms of how easily triggered they are by the inputs and observed by the outputs, respectively.  The states which are neither controllable nor observable or the ones that are weakly controllable or observable are redundant for the input-output behaviour of the system. 
 %A systematic approach to identify a reduced order model by discarding these states is called balanced truncated initially introduced by \cite{moore1981principal} and an approximation is proposed by \cite{rowley2005model} which is referred to as Balanced POD method (BPOD). Balanced truncation amounts to a projection of the system onto the leading balanced modes. Under this transformation and the corresponding adjoint transformation, the controlability and observability Gramians (see \cite{glad2000control}) are identical and diagonal.  
 A limitation pertaining to this method is the necessity of computing the adjoint balanced modes. The Eigensystem Realisation Algorithm (ERA) \cite{juang1985eigensystem,ma2011reduced}  is a system identification technique that allows to circumvent this limitation. It is only based on  sampling measurements extracted directly from the flow, see a detailed description of the method in \cite{ma2011reduced}.

As an example of the performance of the reduced-order model with $r=435$, in Fig.~\ref{fig:mor} we show the impulse responses:
\[
%\phi_{8} \rightarrow {C}_{1,8}{u},\quad w_8 \rightarrow {C}_{2,8}{u}, \quad w_8 \rightarrow {C}_{1,8}{u}.
\phi_{8} \rightarrow {z}_{8},\quad w_8 \rightarrow {y}_{8}{}, \quad w_8 \rightarrow {z}_{8}{}.
\]
In the figure, the solid lines are the impulse response of the full system obtained from solving Navier-Stokes equation while the dotted lines presents the results of the reduced-order model. We observe an equally good agreement for all the inputs and output, when comparing the full system and the ROM.
Now that an efficient ROM is constructed, may design a linear controller.
\subsection{Control Design} 
%The main idea of the linear reactive control is to determine a controller that minimises the energy of disturbances captured by output $C_{1}$. 
We use a classical LQG-approach \cite{lewis,zhou:doyle:glover:02} to determine a controller that minimises the energy of disturbances captured by output $C_{1}$.   
The control signal $\phi(t)$ %should
is designed for the actuator ${B}_2$ such that the mean of the output energy, $z(t)$,
% \begin{align}\label{eq:cost}
% \mathcal{E}(z)={\| z \|^2_{L^2_{[0,\infty]}}} = {\int_0^\infty {u}^T{C}^T_{1}{C}_{1}{u} +\phi^T I_l I_l\phi \hspace{0.2cm} \mathrm{d}t },
% \end{align}
is minimised (see Eq. \ref{eq:cost}).
The LQG design procedure involves a two-step process: first the full state - represented in this case by the velocity field - is reconstructed from the noisy measurement $v(t)$ via an estimator. Once the estimated state $\hat{{u}}$ is computed %based on the noisy measurement ${v}(t)$ 
%
%by marching in time the dynamical system
%\begin{align}
%\dot{\hat{{u}}}(t)={A}_r \hat{{u}}(t) + {B}_{2r} \phi (t)+{L}(C_{2r}\hat{u}-v(t)),
%\label{eq:estimator} 
%\end{align}
%fed by signals extracted from the system. The term ${L} \in \mathbb{R}^{r \times p}$ is the estimator gain and can be computed by solving a Riccati equation \cite{glad2000control}, such that the error $(\hat{u}-u)$ is minimised.
% Once that the state is properly estimated, 
 the control signal can be computed by the following linear relationship
\begin{align}
\phi(t)={K}\hat{{u}}(t),
\label{eq:control_law}
\end{align}
where ${K} \in \mathbb{R}^{m \times r}$ is referred to as the control gain.  When the disturbances are modelled as white Gaussian noise, the separation principle allows the two steps (estimation and full-information control) to be performed independently. Furthermore,  both problems are optimal and stable and the resulting closed loop is also optimal and stable \cite{zhou:doyle:glover:02}. The final form of the reduced order controller (also called compensator) of size $r$ is
\begin{subequations}\label{eqn:aa2}
\begin{alignat}{2} 
\dot{\hat{{u}}}(t)&= ({A}_r+{B}_{2r} {K} + {L} {C}_{2r}) \hat{{u}}(t) - {L} {v}(t), \\
\phi(t)&={K} \hat{{u}}(t),
\end{alignat}
\end{subequations}
where the term ${L} \in \mathbb{R}^{r \times p}$ is the estimator gain and can be computed by solving a Riccati equation \cite{glad:ljung}, such that the error $\epsilon=\|\hat{u}-u\|^2$ is minimised. The controller is thus a state-space system with the measurements $v(t)$ as input and the control signal $\phi(t)$ as output.
%
%Integrating the compensator with the full Navier-Stokes equations yields the closed-loop system
%\begin{align}
%\left(
%\begin{array}{c}
%\dot{{u}} \\
%\dot{\hat{{u}}}
%\end{array}
%\right) = 
%\left(
%\begin{array}{cc}
%{A} & {B}_{2}{K}\\
%-LC_2 & {A}_r+{B}_{2r}{K}+{L}{C}_{2r}
%\end{array}
%\right)
%\left(
%\begin{array}{c}
%{u} \\
%\hat{{u}}
%\end{array}
%\right)+  
%\left(
%\begin{array}{cc}
%{B}_{1} & 0\\
%0 & -{L}
%\end{array}
%\right)
%\left(
%\begin{array}{c}
%w \\
%I_{\alpha}g
%\end{array}
%\right).
%\label{equ:CL}
%\end{align}
The evolution of the perturbations is simulated by marching in time the full DNS, while the controller runs on-line, simultaneously. Eq.~\ref{eqn:aa2}$a$ is based on the reduced-order model and is solved by using a standard Crank-Nicholson scheme.
%The spatio-temporal evolution of the disturbances governing by the closed loop system is computed by solving simultaneously the full DNS together with the reduced-order estimation. %The calculation of states ${u}$ is performed using a timestepper (\texttt{nek5000}) and 

% 
% 
% 
% \begin{figure}%[t!]
% \begin{center}
% \includegraphics[width=0.77\textwidth]{./pic/cent/mor-b2-8-c1-8.eps} 
% % \put(-305,0){{\begin{sideways}{${B}_{2,8} w \rightarrow {C}_{1,8}{u}$}  \end{sideways}}} 
% \put(-305,0){(a)} \\
% \includegraphics[width=0.77\textwidth]{./pic/cent/mor-b1-8-c2-8.eps} 
% % \put(-305,0){{\begin{sideways}{${B}_{1,8} w \rightarrow {C}_{2,8}{u}$}  \end{sideways}}} 
% \put(-280,40){(b)} \\
% \includegraphics[width=0.77\textwidth]{./pic/cent/mor-b1-8-c1-8.eps} 
% % \put(-305,0){{\begin{sideways}{${B}_{1,8} w \rightarrow {C}_{1,8}{u}$}  \end{sideways}}} 
% \put(-305,0){(c)} 
% \put(-280,-0){$x$}
% \end{center}
% \caption{Impulse response of the system from the input $B_{2,8}$ to the output $C_{1,8}$ (a), from the input $B_{1,8}$ to the output $C_{2,8}$ (b) and  from the input $B_{1,8}$ to the output $C_{1,8}$ (c). The red line shows the DNS results, while the dotted lines indicates the impulse response of the reduced model (Case N Tab. \ref{tab:pf}) }
% \label{fig:mor}
% \end{figure}
\subsection{Centralised and Decentralised Controllers}
A multivariable control approach is necessary since our system has more than one actuator and sensor.
The degree of control complexity in a multivariable approach depends on the degree of coupling between inputs and outputs. For example consider the transfer function between the input $w_j$ to the output $y_k$. Then the effect on $y_k$ due to a small change in $w_j$ may depend on one, a few or all other inputs $w_l$ for $l\neq j$,  if the system is uncoupled, weakly coupled or fully coupled, respectively. The degree of coupling  depends usually not only on the actuator/sensor placement but also on the dynamics of the TS wavepackets. As we shall see, we have a situation of a weakly coupled system, due to the fact that a TS wavepacket generated from a point source spreads only in a limited spanwise region.

The most straight-forward approach is the so called centralised controller where all the inputs and outputs are connected together. The main disadvantage is that the number of interconnections -- thus  the complexity of the controller -- increase significantly as we aim to control perturbations over a larger span of the domain. In contrast, a fully decentralised controller connects only one sensor to one actuator, and thus requires by definition the same number of actuators and sensors. This approach disregards any influence of an input which is not placed directly upstream the output; this is a risky model assumption, as the influence that may exist in reality will induce an over- or underestimation of the signals, causing instabilities. A compromise between the centralized and fully decentralized approach is a semi-decentralized approach (henceforth only referred to as decentralized), 
where the system is divided into a collection of independent sub-systems. For each sub-system a controller is designed -- called a control unit -- for a few number of sensors and actuators. Then, the same controller is replicated along the span to cover a broader region.
 As we will see the division into control units provides an efficient means for control of TS waves, since the disturbance source upstream is only observable at a subset of sensors; thus some of the interconnections which are not relevant to the dynamics of the system are neglected (see Fig.~\ref{fig:config}). 
%%%%%%%%%%%%%%%%%%%%%%%%%%%%%%%%%%%%%%%%%%%%%%%%%%%%%%%%%%%%%%%%%%%%%%%%%%%%%%%%%%%%%%%%
%                              Result
%%%%%%%%%%%%%%%%%%%%%%%%%%%%%%%%%%%%%%%%%%%%%%%%%%%%%%%%%%%%%%%%%%%%%%%%%%%%%%%%%%%%%%%%%
\section{Results}
In the following sections, we first design and analyse a centralised controller for the  attenuaton of small--amplitude  TS wavepackets. After a parametric study of the control penalty, we identify a reference controller, as the centralized controller that for the chosen flow parameters ($Re$, domain, etc) provides the best performance. Second, we design a set of decentralised controllers by assembling    several control units of different sizes. Their control efficiency in terms of performance (robustness is left for future studies) will be compared to the reference controller.

%------------------------------------------ table 18-18-18-18--------------------
% \begin{table}
% \begin{tabular}{c c c c c c}
% & & & & & \\ % put some space after the caption
% \hline
% \hline
% Case  & Description & $\text{Cont. penalty}$ & $\text{Model order}$ & $\text{Norm}$ & $\text{Eng. reduction}$\\
%  $ k $& $ - $       &$l$ & $r$ & $1-\frac{\parallel G_{k} \parallel_2^2}{\parallel G_{n} \parallel_2^2}$ & $\bar{E_r}$\\
% 
% \hline
% N   & $18/18-18-18-18/1$ &$ - $ &$ - $&$0\%$&$ - $ \\
% A   & $18/18-18-18-18/1$ &$100$ &$435$&$45\%$ &$0.27$\\
% B   & $18/18-18-18-18/1$ &$10 $ &$435$&$98\%$ &$0.80$\\
% C   & $18/18-18-18-18/1$ &$1  $ &$435$&$98\%$ &$0.80$\\
% \hline
% \hline
% \end{tabular}
% \centering

\begin{table}
\begin{tabular}{p{1cm}  p{4cm} p{1cm} p{1cm} p{1.5cm} p{1cm} }
% \begin{tabular}{c c c c c c}
& & & & & \\ % put some space after the caption
\hline
\hline
Case  & Description & $\text{Control}$ & $\text{Order}$ & $\text{Norm}$ & $\text{Energy}$\\
      &             & $\text{Penalty}$ &                &$\text{Reduction}$ & $\text{Reduction}$\\
 $ k $& $ - $       &$l$ & $r$ & $1-\frac{\parallel G_{k} \parallel_2^2}{\parallel G_{n} \parallel_2^2}$ & $E_r$\\

\hline
N   & $18/18-18-18-18/1$ &$ - $ &$ - $&$0\%$&$ - $ \\
A   & $18/18-18-18-18/1$ &$100$ &$435$&$45\%$ &$0.27$\\
B   & $18/18-18-18-18/1$ &$10 $ &$435$&$98\%$ &$0.80$\\
C   & $18/18-18-18-18/1$ &$1  $ &$435$&$98\%$ &$0.80$\\
\hline
\hline
\end{tabular}
\centering
\caption{The performance of a LQG controller designed with different control penalties. The noise autocovariance on the estimation sensors and for all cases are assumed constant $\alpha^2 = 10^{-6}$. The norms are computed in the time interval $t \in [2000, 8000]$. The description identifier is defined as the following; number of disturbances $B_1$ $/$ the design configuration of the system consists of $d-p-m-k$ disturbances-estimation sensors-actuators-outputs$/$ number of control units.} 
\label{tab:pf}
\end{table}
%%%%%%%%%%%%%%%%%%%%%%%%%%%%%%%%%%%%%%%%%%%%%%%%%%%%%%%%%%%%%%%%%%%%%%%%%%%%%%%%%%%%%%%%%%%%%%%
\subsection{Centralised  Controller}
%%%%%%%%%%%%%%%%%%%%%%%%%%%%%%%%%%%%%%%%%%%%%%%%%%%%%%%%%%%%%%%%%%%%%%%%%%%%%%%%%%%%%%%%%%%%%%%
In Tab. \ref{tab:pf} the effect of different control penalties (parameter $l$ �in Eq. \ref{eq:cost}) on the performance of the closed-loop system is investigated for a centralized LQG controller and the setup in Tab. \ref{tab:elements}. The optimal value of the control penalty  is usually not known before applying the controller to the full DNS and involves an iterative procedure. In general, small values of the control penalty correspond to a reduction of the perturbation amplitude; however, too low values of control penalties result in unfavourable behaviour such as spurious control signal. Case $C$ in Tab. \ref{tab:pf} is selected as the baseline reference controller, for which all decentralized controller will be compared to.

\begin{figure}%[t!]
\begin{center}
\includegraphics[width=0.88\textwidth]{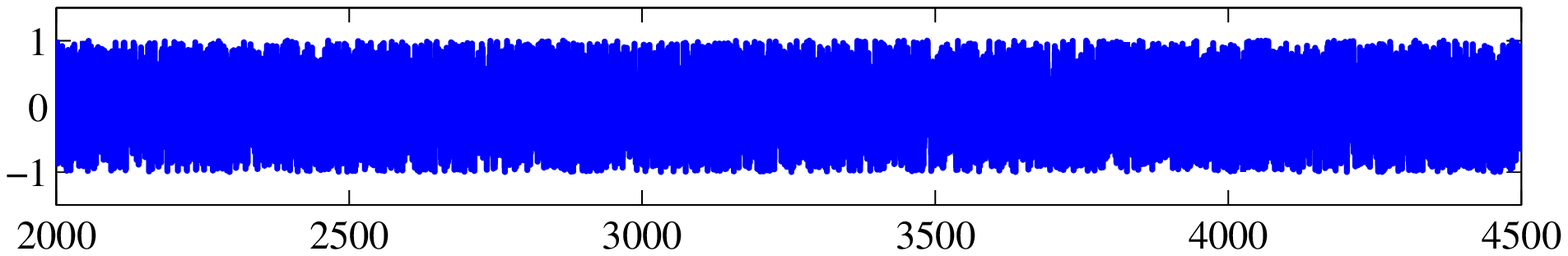} 
\put(-310,22){{\begin{sideways}{$w_8$}  \end{sideways}}} 
\put(-330,45){(a)} \\
\includegraphics[width=0.88\textwidth]{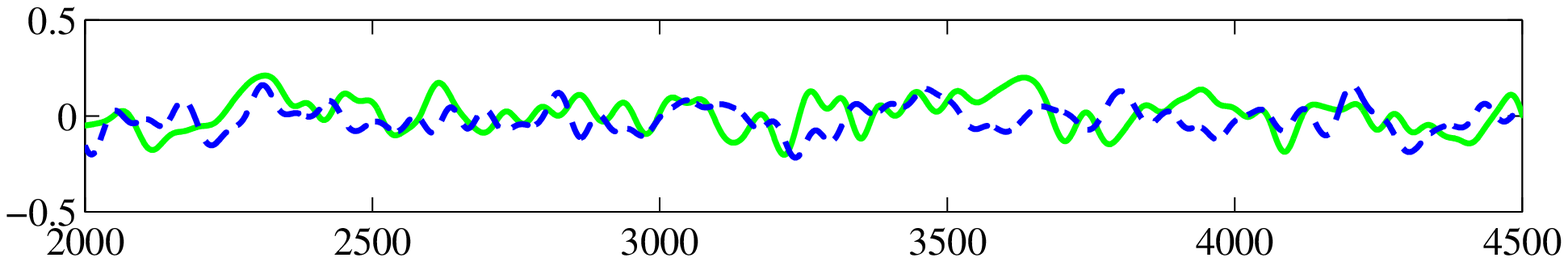}  
\put(-310,10){{\begin{sideways}{$w \rightarrow {C}_2{u}$}  \end{sideways}}} 
\put(-330,45){(b)} \\
\includegraphics[width=0.88\textwidth]{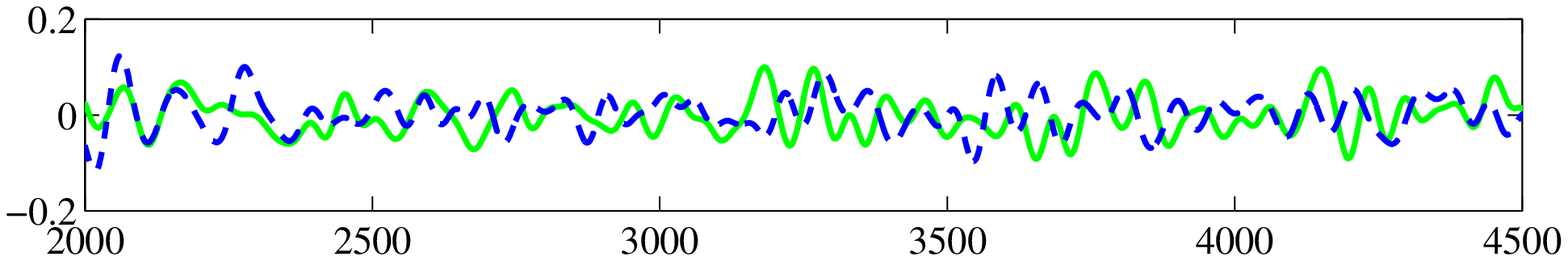}  
\put(-310,20){{\begin{sideways}{$\phi$}  \end{sideways}}} 
\put(-330,45){(c)} \\
\includegraphics[width=0.88\textwidth]{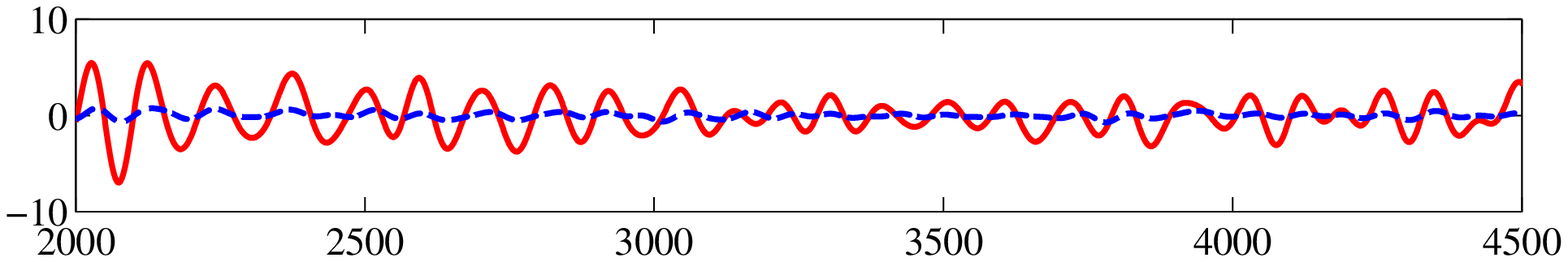}  
\put(-310,10){{\begin{sideways}{$w \rightarrow {C}_{1,8} {u}$}  \end{sideways}}} 
\put(-330,45){(d)} 
\put(-160,-10){$t$}   
\end{center}
% \begin{figure}%[t!]
% \begin{center}
% \includegraphics[width=0.88\textwidth]{./pic/cent/fb-signal-b1.eps} 
% \put(-350,30){{\begin{sideways}{$w_8$}  \end{sideways}}} 
% \put(-350,60){(a)} \\
% \includegraphics[width=0.88\textwidth]{./pic/cent/fb-signal-c2.eps}  
% \put(-350,10){{\begin{sideways}{$w \rightarrow {C}_2{u}$}  \end{sideways}}} 
% \put(-350,60){(b)} \\
% \includegraphics[width=0.88\textwidth]{./pic/cent/fb-signal-ct.eps}  
% \put(-350,30){{\begin{sideways}{$\phi$}  \end{sideways}}} 
% \put(-350,60){(c)} \\
% % \hspace{0.3 cm}
% \includegraphics[width=0.88\textwidth]{./pic/cent/fb-signal-c18.eps}  
% \put(-350,10){{\begin{sideways}{$w \rightarrow {C}_{1,8} {u}$}  \end{sideways}}} 
% \put(-350,60){(d)} 
% \put(-180,-10){$t$}   
% \end{center}
\caption{Noise response of the closed-loop system: Stochastic excitation of the input ${B}_{1,8}$ is shown in (a),  estimation signals $C_{2,8}$ (dashed blue line) and  $C_{2,18}$ (solid green line) in (b), control signal feeding the actuator ${B}_{2,8}$ (dashed blue lines) and ${B}_{2,18}$ (solid green line) in ($c)$ and measurement extracted by sensor ${C}_{1,8}$ for uncontrolled (solid line) and controlled and dashed (dashed line) system (cases N and C in Tab. \ref{tab:pf}) in (d).}
\label{fig:fb_signal}
\end{figure}

First, we characterize the performance of controller C using a number of different observables.  Fig.~\ref{fig:fb_signal} represents the input-output behaviour of the closed-loop system for case $C$.  In this setup, there are totally $18$ inputs $B_1$;  each of them are exited by an independent white noise of variance $\frac{1}{3}$. In the first frame (Fig.~\ref{fig:fb_signal}a), the disturbance input $w_{8}$ is shown. It is a white noise signal that provides a continuous forcing at $B_{1,8}$. Fig.~\ref{fig:fb_signal}b shows the measurement detected by upstream sensors $C_{2,8}$ and $C_{2,18}$. The sensors are located close to the wall, inside of the boundary layer and can register the evolution of the disturbance. One clearly observes that certain frequencies are amplified by the system, whereas others are damped. 
Fig.~\ref{fig:fb_signal}c reports the control signals related to actuators $B_{2,8}$ and $B_{2,18}$. Since the disturbances are uncorrelated, we can observe independent behaviour for different actuators. Finally, in Fig.~\ref{fig:fb_signal}d, the signal extracted from output $C_{1,8}$ for the uncontrolled and controlled cases is shown. The root mean square (r.m.s) of the signal is reduced up to $89\%$.  
%------------------------------------------ figure 18-18-18-18 performance-------------------
%------------------------------------------ figure 18-18-18-18 performance-------------------

\begin{figure}%[t!]
\centering
\includegraphics[width=0.69\textwidth]{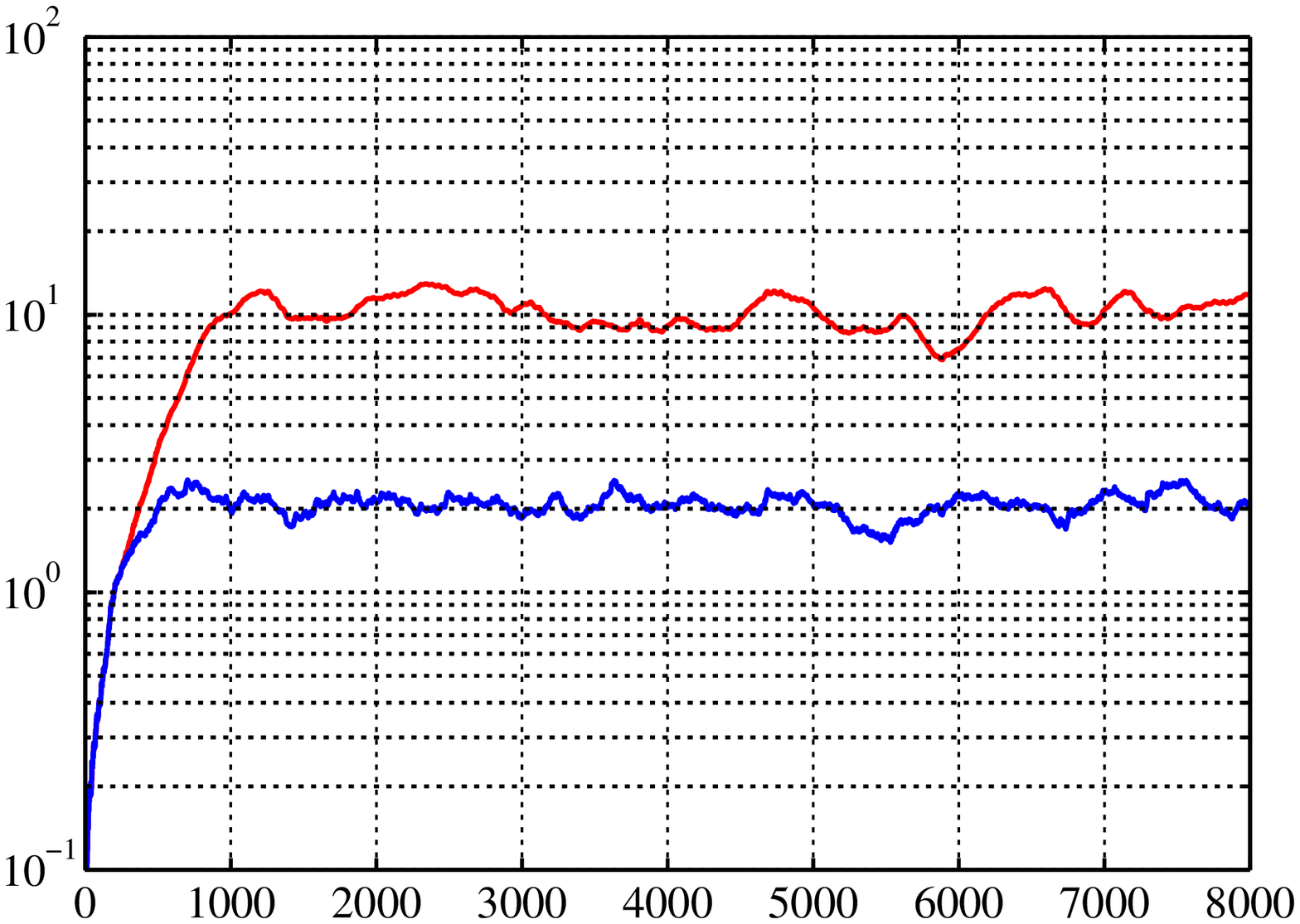} 
\put(-240,80){$E$}
\put(-100,0){$t$}
\caption{Energy of the perturbations $E$ as a function of time $t$; solid red line corresponds to the energy of uncontrolled case $E_n$ and solid blue line to the controlled case $E_c$. The statistics are computed for the time interval $t \in [3000 \hspace{0.2 cm} 8000]$.}
\label{fig:cent_eng}
\end{figure}

In a three-dimensional configuration, the minimisation of the sensor measurements near the wall, does not guarantee the reduction of the perturbation energy in the full domain. This has to be evaluated a posteriori. Fig.~\ref{fig:cent_eng} shows the energy, $E(t) = u^Tu/2$, of the perturbation as a function of time.
% It is defined as
% \begin{equation}
%  {E} = \frac{1}{2}\int_\Omega{u^Tu}\,{dv}.
% \end{equation}
%\begin{equation}
% {E(t)} = \frac{1}{2}\|u\|_2^2
 % {E(t)} = \frac{1}{2}u^Tu
%\end{equation}
%
%where $u$ is the velocity vector of the disturbances when the system is excited with a stochastic white-noise forcing. 
The mean value of the energy reduction $\bar{E_r}$ is defined as
\begin{equation}
 \bar{E_r}=\frac{\int_{t_0}^{t_1}{{E}_n}dt -\int_{t_0}^{t_1}{{E}_c}dt}{\int_{t_0}^{t_1}{E}_n dt},
\end{equation}
where  $[t_0,\hspace{0.1cm}t_1]$ is the time interval in which the statistics are computed. In Fig.~\ref{fig:cent_eng}, the uncontrolled energy $E_n$ is shown by a solid red line while the controlled energy, $E_c$ is shown with a blue line. We observed that the energy is reduced by approximately $80\%$.  
%------------------------------------------ figure 18-18-18-18 energy-------------------
%------------------------------------------ end figure 18-18-18-18energy-------------------

\begin{figure}%[t!]
\centering
\includegraphics[width=0.49\textwidth]{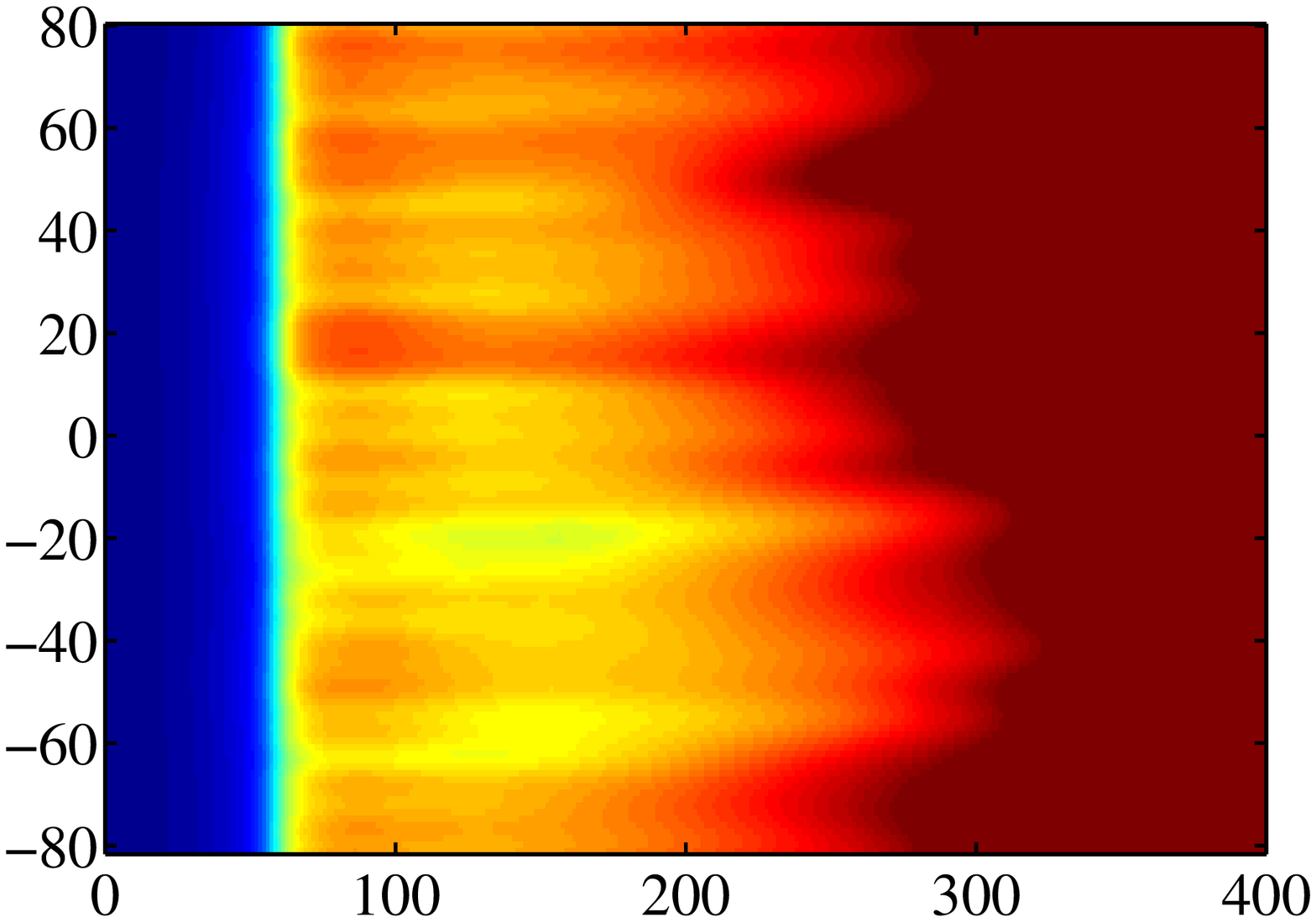} 
\put(-90,-10){$x$}
\put(-180,55){$z$}
\put(-180,110){(a)}
\includegraphics[width=0.49\textwidth]{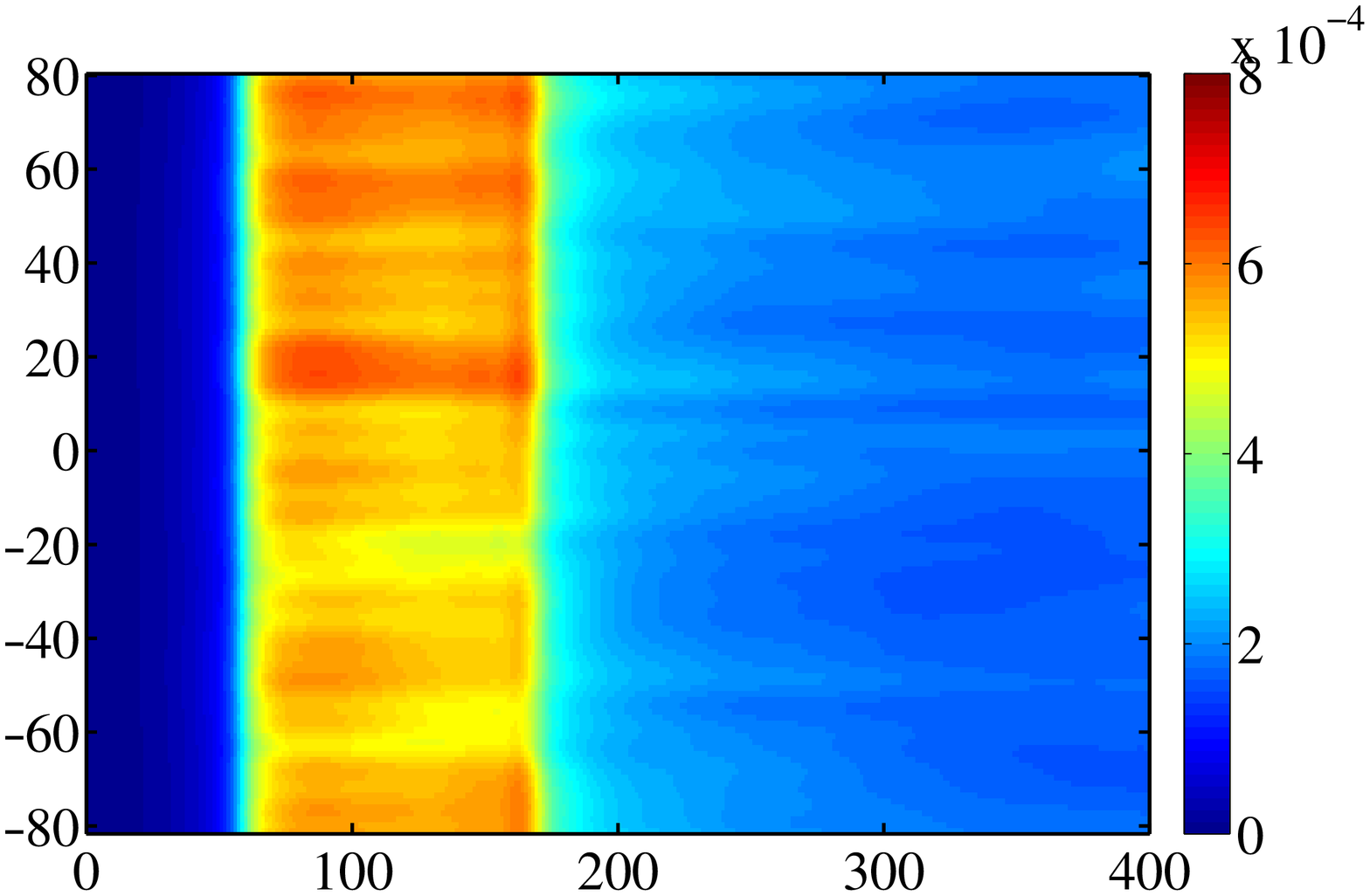} 
\put(-100,-10){$x$}
\put(-180,110){(b)}\\
\includegraphics[width=0.49\textwidth ]{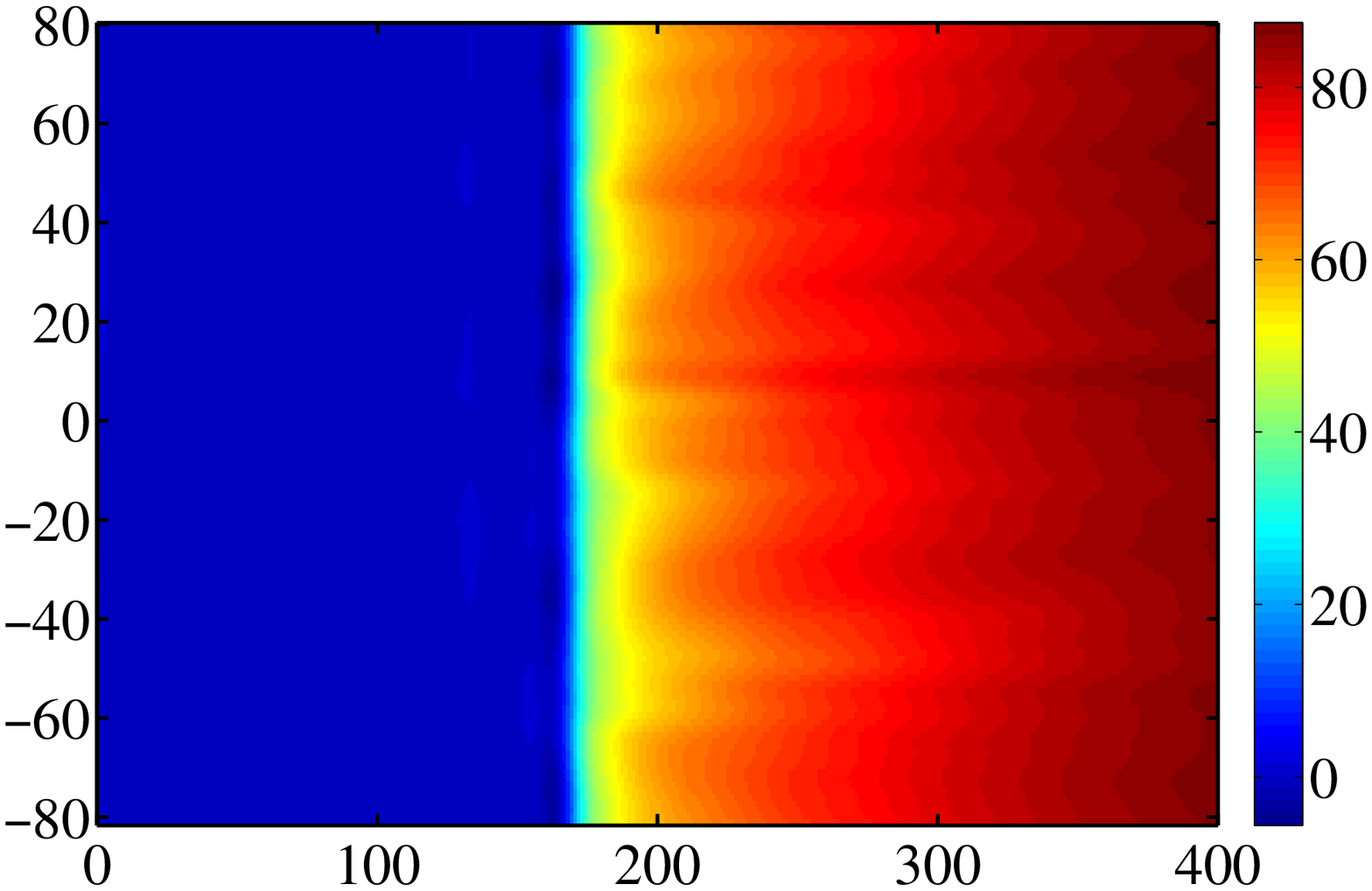} 
\put(-90,-10){$x$}
\put(-180,55){$z$}
 \put(-200,120){(c)}
\caption{Streamwise root mean square velocity averaged along wall normal direction for the uncontrolled case N (a) and  controlled case C (b) and the corresponding percentage of the reduction (c). The statistics are computed for the time interval $t \in [3000, \hspace{0.2 cm} 8000]$}
\label{fig:cent_spatial}
\end{figure}

Finally, in order to gain an insight into where in the physical domain, the controller has a strong effect, 
we show in Fig.~\ref{fig:cent_spatial}  the distribution of the r.m.s value of the streamwise velocity of disturbances in horizontal plane (streamwise-spanwise)
%  For any vector $v \in \mathbb{R}^n$, r.m.s is defined as
% \begin{equation}
%  v_{rms}=\sqrt{\frac{1}{T}\int_0^T{v^2}{dt} - (\frac{1}{T}\int_0^T{v}{dt})^2};
% \end{equation}
% The r.m.s of the velocities are 
averaged along wall normal direction. The disturbances  $B_1$ are located at $x=60$ from the beginning of the computational box. We expect the amplitude of the perturbations to grow as we move toward the end of the domain in uncontrolled case N (Fig.~\ref{fig:cent_spatial}a). Fig.~\ref{fig:cent_spatial}b shows the resulting r.m.s value of the perturbations when the controller is active. The suppression of the perturbations begin from $ x=167 $  where the actuators are located. Fig.~\ref{fig:cent_spatial}c reports the percentage of the reduction in r.m.s of the perturbation. Since the objective function of the controller is to attenuate the amplitude of the perturbation where the outputs are located, a significant reduction is observed at that region; the reduction is also homogeneous in spanwise direction. %This is related to the distance between the actuator. In fact, if the distance increases too much we cannot maintain the homogeneity anymore in this direction.
%------------------------------------------ figure 18-18-18-18 rms----------------
\subsection{Decentralised Controllers}
%%%%%%%%%%%%%%%%%%%%%%%%%%%%%%%%%%%%%%%%%%%%%%%%%%%%%%%%%%%%%%%%%%%%%%%%%%%%%%%%%%%%%%%%%%%%%%%
% The centralised controller discussed in previous section can successfully attenuate the amplitude of the disturbance in the entire region and in particular along the total span of the domain. However, as we require to cover a larger area, it is demanded to significantly increase the complexity of the controller; in fact, all the sensors and actuators are connected together and the number of interconnections increase rapidly so does the complexity of the system. One solution to circumvent this restriction is to used decentralised controller where the domain is divided into sub-systems and for each sub-system a a control unit is designed with limited complexity. 
Having shown that centralized controller with a very high complexity may reduce energy by nearly an order of magnitude, we now investigate how decentralized  controllers of lower complexity compare in performance. As already mentioned, the decentralized controllers are designed in two steps; (i) constructing a control unit using only a few actuators and sensors; (ii) by replicating the units in the spanwise direction. 

%In the following sections we investigate the performance of different types of control units when only a single control unit is active and   when we replicate the controllers along the span. Since a single control unit is designed to attenuate the amplitude of a restricted number of disturbances, we expect only to reduce the disturbances locally by a small amount.       
\begin{figure}%[h!]
\begin{center}
\includegraphics[width=0.48\textwidth]{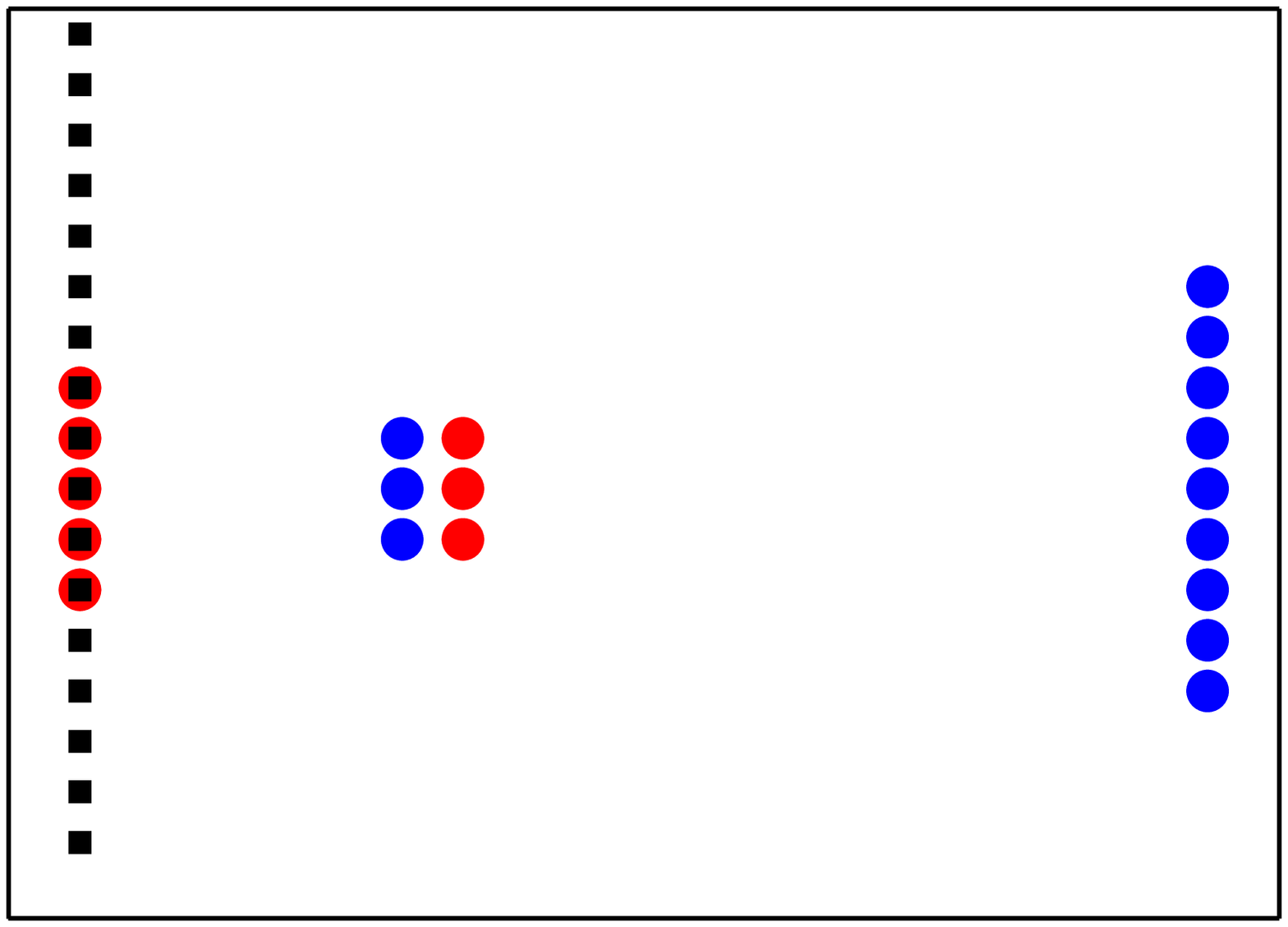}  
\put(-165,125){(a)}
%\put(-120,100){\tiny asymmetric control unit}
\put(-100,25){\tiny 18/5-3-3-9/1}
\includegraphics[width=0.48\textwidth]{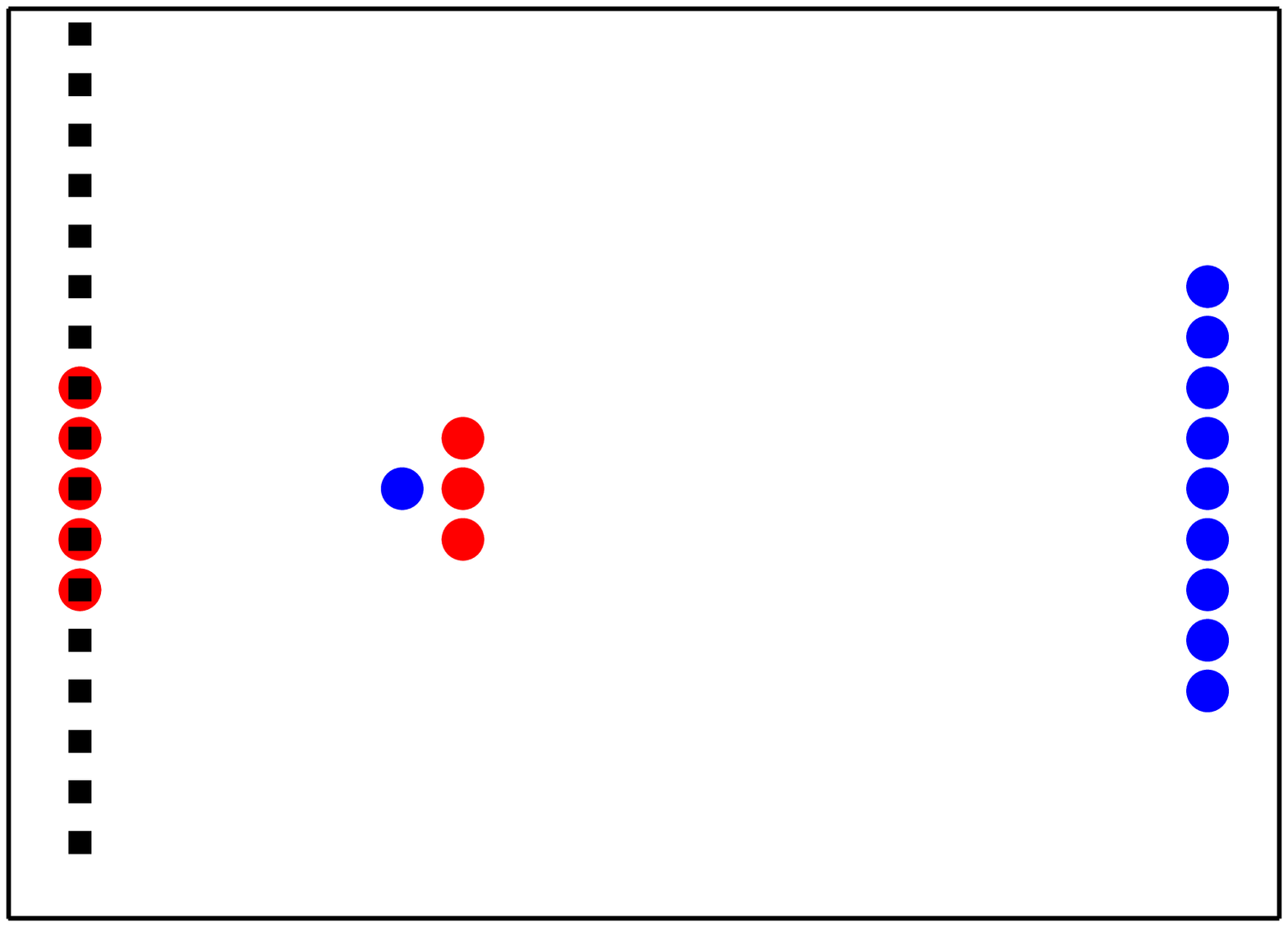}  
\put(-165,125){(b)}  
%\put(-120,100){\tiny symmetric control unit}  
\put(-100,25){\tiny 18/5-1-3-9/1}
\end{center}
\caption{A schematic view of two control units. The controller shown in (a) is designed considering $5$ upstream disturbances $B_{1,(6,7,\cdots,10)}$, $3$ estimation sensors $C_{2,(7,8,9)}$, $3$ actuators $B_{2,(7,8,9)}$ and $9$ outputs $C_{1,(4,5,\cdots, 12)}$ as the objective function (circles). This control unit performs when $18$ disturbances are evolving into the domain (squares). The layout and the number of sensors and actuators remain the same for the control unit depicted (b), but only one estimation sensor $C_{2,8}$ is used.}
\label{fig:sketch_sa}
\end{figure} 

%%%%%%%%%%%%%%%%%%%%%%%%%%%%%%%%%%%%%%%%%%%%%%%%%%%%%%%%%%%%%%%%%%%%%%%%%%%%%%%%%%%%%%%%%%%%%%
\subsubsection{Design and Performance of Single Control Units}
%%%%%%%%%%%%%%%%%%%%%%%%%%%%%%%%%%%%%%%%%%%%%%%%%%%%%%%%%%%%%%%%%%%%%%%%%%%%%%%%%%%%%%%%%%%%%%%

\begin{table}
\begin{tabular}{p{1cm}  p{3cm} p{1cm} p{1cm} p{1.5cm} p{1cm} }
% \begin{tabular}{c c c c c c}
& & & & & \\ % put some space after the caption
\hline
\hline
Case  & Description & $\text{Control}$ & $\text{Order}$ & $\text{Norm}$ & $\text{Energy}$\\
  &  & $\text{penalty}$ & $\text{ }$ & $\text{ Reduction}$ & $\text{Reduction}$\\
 $ k $& $ - $       &$l$ & $r$ & $1-\frac{\parallel G_{k} \parallel_2^2}{\parallel G_{n} \parallel_2^2}$ & $\bar{E_r}$\\

\hline
D   & $18/5-3-3-9/1$ &$20 $ &$155$ &$4.6\%$&  $0.109$\\[3pt]
E   & $18/5-1-3-9/1$ &$20 $ &$155$ &$2.2\%$&  $0.044$\\
F   & $18/3-3-3-9/1$ &$20 $ &$119$ &$3.4\%$&  $0.087$\\
G   & $18/5-3-3-3/1$ &$10 $ &$87$  &$8.4\%$&  $0.083$\\
\hline
\hline
\end{tabular}
\centering
\caption{In each case only one control unit is employed. The noise autocovariance for all the cases are assumed as $\alpha^2 = 10^{-6}$ and the norms are computed for time $t > 2000$.}
\label{tab:cont_unit}
\end{table}

Motivated by the experimental work of  \cite{2006:li:gaster}, we choose to investigate two control units:
\begin{enumerate}
\item The first one consists of three actuators (the center actuator $B_{2,8}$ and two adjacent to the center $B_{2,7}$ and $B_{2,9}$), three estimation sensors ($C_{2,7},C_{2,8}$ and $C_{2,9}$) and $9$ objective sensors $C_{1,(4,5,\cdots, 12)}$. 
During the design process of the control unit, we assume that there exists $5$ upstream disturbances $B_{1,(6,7,\cdots,10)}$, but the actual performance of the controller is assessed when $18$ disturbance sources are active (see sketch in Fig.~\ref{fig:sketch_sa}$a$).  The description identifier of this control unit is ($18/5-3-3-9/1$), where the different numbers are respectively; number of disturbances B1 / the design configuration of the system consists of $d$ - $p$ - $m$ - $k$ (disturbances-estimation sensors-actuators-outputs) / number of control units. 

\item The second one ($18/5-1-3-9/1$) has  only one estimation sensor, namely the center one ($C_{2,8})$ as shown in Fig.~\ref{fig:sketch_sa}$b$. The remaining parameters are the same the first control unit.

\end{enumerate}

\begin{figure}%[t!]
\begin{center}
\includegraphics[width=0.88\textwidth]{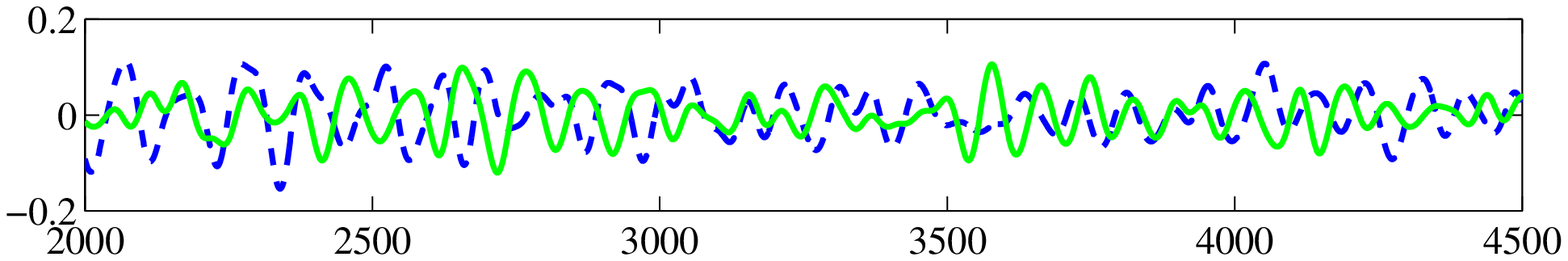}  
\put(-310,23){{\begin{sideways}{$\phi$}  \end{sideways}}} 
\put(-310,50){(a)} \\
\includegraphics[width=0.88\textwidth]{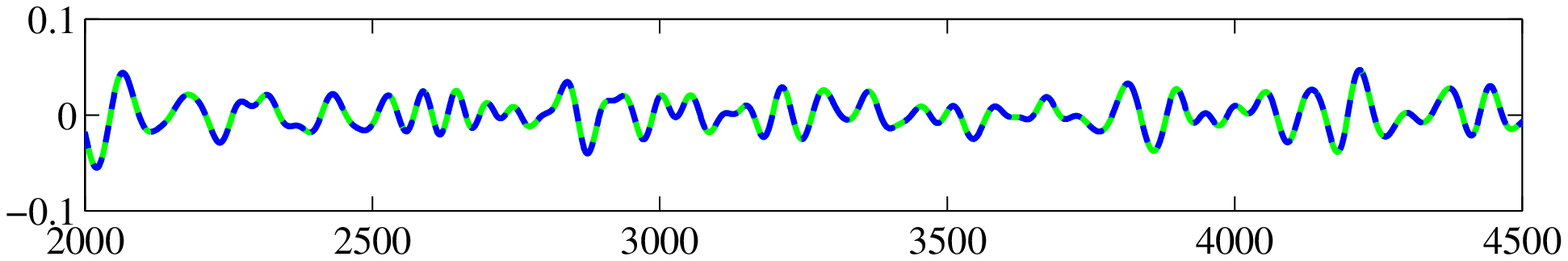}  
\put(-310,23){{\begin{sideways}{$\phi$}  \end{sideways}}} 
\put(-310,50){(b)}
\put(-150,-10){t}   
\end{center}
\caption{Control signal driving the actuators $B_{2,7}$ (solid line) and  $B_{2,9}$ (dotted line) are shown in (a) for a three-estimation sensors-based control unit (case $D$ in Tab. \ref{tab:cont_unit}) and in (b) for single-estimation sensors-control unit (case $E$ in Tab. \ref{tab:cont_unit}).}
\label{fig:fb_decent}
\end{figure}

\begin{figure}%[t!]
\centering
\includegraphics[width=0.49\textwidth]{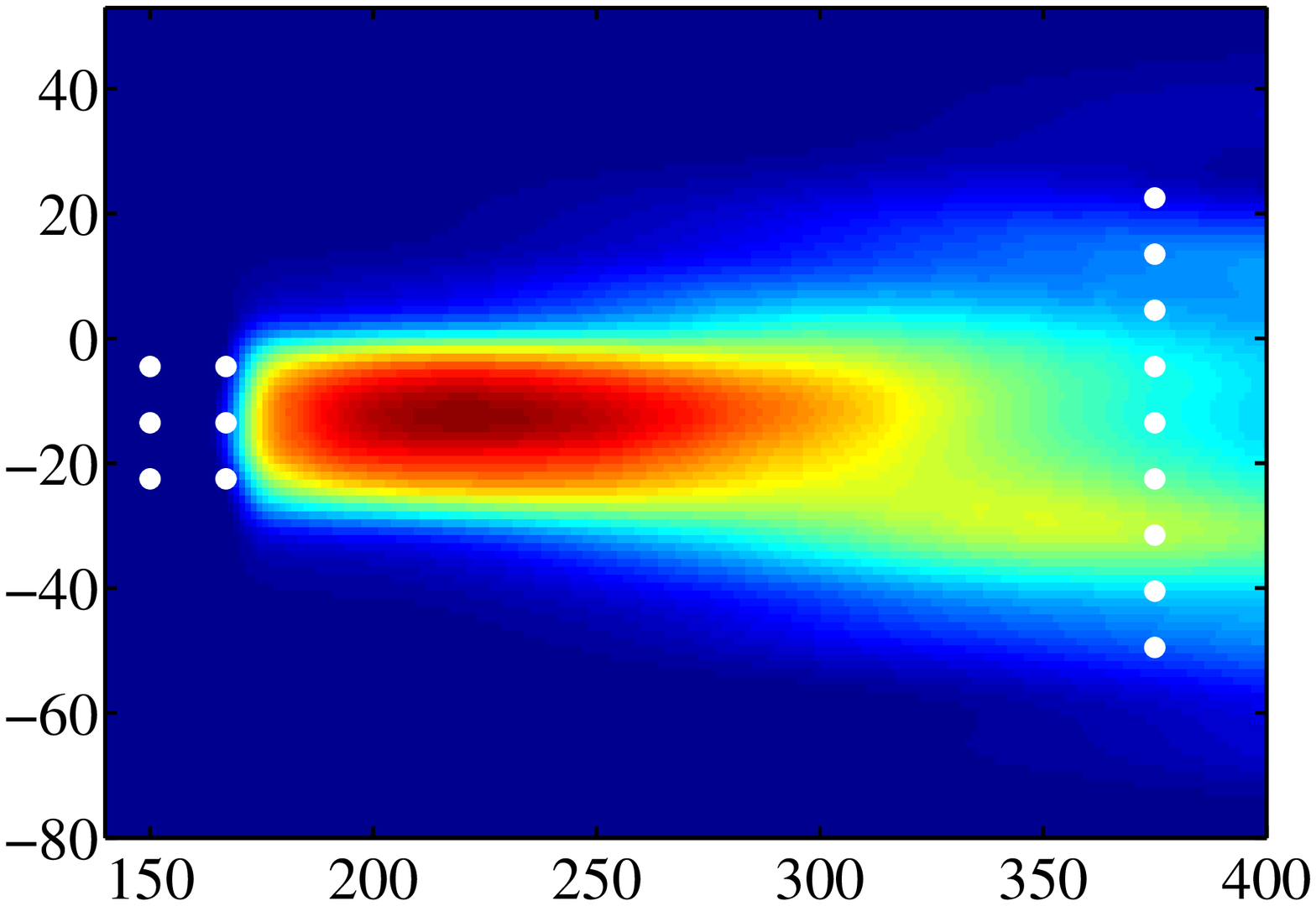} 
\put(-90,-5){$x$}
\put(-180,55){$z$}
\put(-180,110){(a)}
\includegraphics[width=0.49\textwidth]{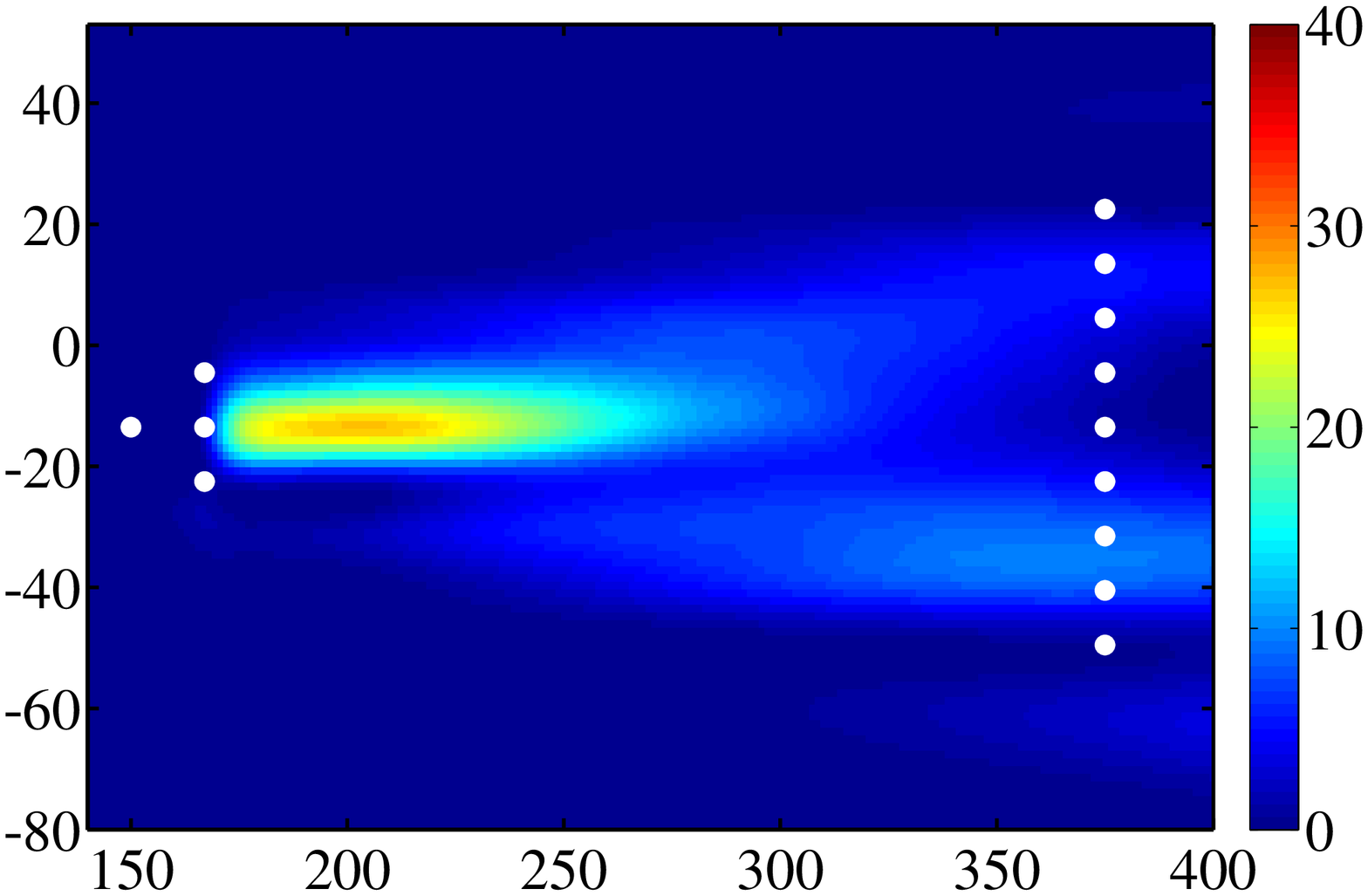} 
\put(-90,-5){$x$}
\put(-180,110){(b)} 
\caption{Percentage reduction in streamwise velocity cancellation averaged along wall normal direction for case $D$ (a) and $E$ (b) is shown. The white dots indicate the location of sensors $C_1$, $C_2$ and actuators $B_2$.}
\label{fig:fb1}
\end{figure}

% After designing the control units, their performances are monitored while $18$ disturbances $B_1$ evolve and convect downstream ($18/5-3-3-9/1$).
 Fig \ref{fig:fb_decent} shows the control signal for the two lateral actuators $B_{2,7}$ and $B_{2,9}$ for both control unit one and two. It is obvious that the two actuators behave in the same manner for the second controller (case $E$ in Tab. \ref{tab:cont_unit}) while they are acting independently for the multiple sensor control unit (case $D$ in Tab. \ref{tab:cont_unit}).
%The performance of a single asymmetric and symmetric control unit are depicted in Tab. \ref{tab:cont_unit} (case $D$ and $E$) and a schematic view is presented in Fig.~\ref{fig:sketch_sa}. In this paper, we refer to a control unit as a symmetric one when the computed control signals driving the actuators $B_{2,7}$ and $B_{2,9}$, situated symmetrically on the opposite sides of the central actuator $B_{2,8}$ are the same and asymmetric otherwise (see Fig.~\ref{fig:fb_decent}). The asymmetric controller is designed considering $5$ upstream disturbances $B_{1,(6,7,\cdots,10)}$ evolving inside the boundary layer; $3$ estimation sensors $C_{2,(7,8,9)}$ and $3$ actuators $B_{2,(7,8,9)}$ responsible to identify and quench the upcoming perturbations and $9$ outputs $C_{1,(4,5,\cdots, 12)}$ as the objective function downstream of the domain employed to assess the performance of the controller. The layout and the number of sensors and actuators remain the same for the symmetric control unit; however we used only one estimation sensor $C_{2,8}$ upstream of the actuators ($5-3-3-9/1$). After designing the control units, their performances are monitored while $18$ disturbances $B_1$ evolve and convect downstream ($18/5-3-3-9/1$). Fig \ref{fig:fb_decent} depicts the control signal for the two lateral actuators $B_{2,7}$ and $B_{2,9}$ for both cases. It is obvious that the two actuators behave in the same manner for the symmetric controller (case $E$) while they are acting independently for the asymmetric one (case $D$).
%------------------------------ figure control signal  -------------
%------------------------------ end figure control signals -------------
%
Fig.~\ref{fig:fb1} shows the streamwise velocity cancellation averaged along wall normal direction. The white dots indicate  the spatial configuration of the sensors and actuators for the two cases $D$ and $E$. The Figs.~\ref{fig:fb1}a and \ref{fig:fb1}b confirm that a level of cancellation up to $40\%$ is achieved in the central area adjacent to downstream of the actuators while it faded away as we move downstream. Controller based on only one upstream sensor can act on a limited region while the controller based on three sensors is able to influence a broader domain. The reason is that the latter controller can identify the discrepancy between the disturbances coming from lateral sides, i.e. the observability of the system is significantly larger. This controller can attenuate the energy of the system up to $10.9\%$ (\emph{see} Tab.~\ref{tab:cont_unit} case $D$), while the single-sensor controller can only suppress the energy up to $4.4\%$. Furthermore, in terms of norms of the objective function, the corresponding reduction between the two controllers are  $4.6\%$  and $2.2\%$. In the following section we use the first control unit in the list above.

%------------------------------ figure control unit performance  -------------
%------------------------------ end figure control unit performance  -------------

%------------------------------  figure perturbation energy  -------------
\begin{figure}%[t!]
\centering
\includegraphics[width=0.49\textwidth]{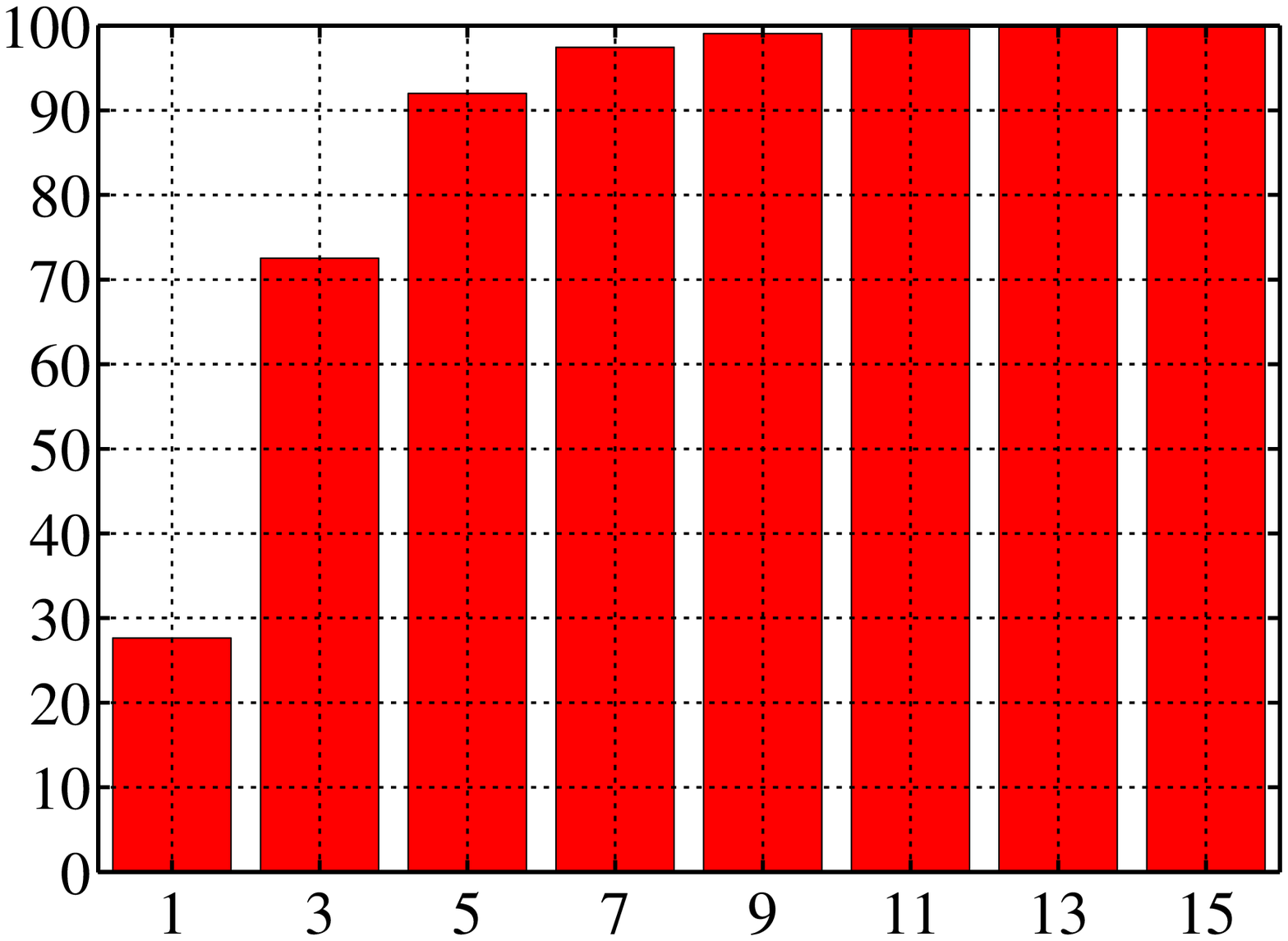}
%\setlenght{\textwidth}
\put(-170,40){{\begin{sideways}{$E_{B_{1,j}\rightarrow C_{2,(7,8,9)} \%}$}  \end{sideways}}} 
\put(-80,0){$j$}
\put(-170,120){(a)} 
\includegraphics[width=0.49\textwidth]{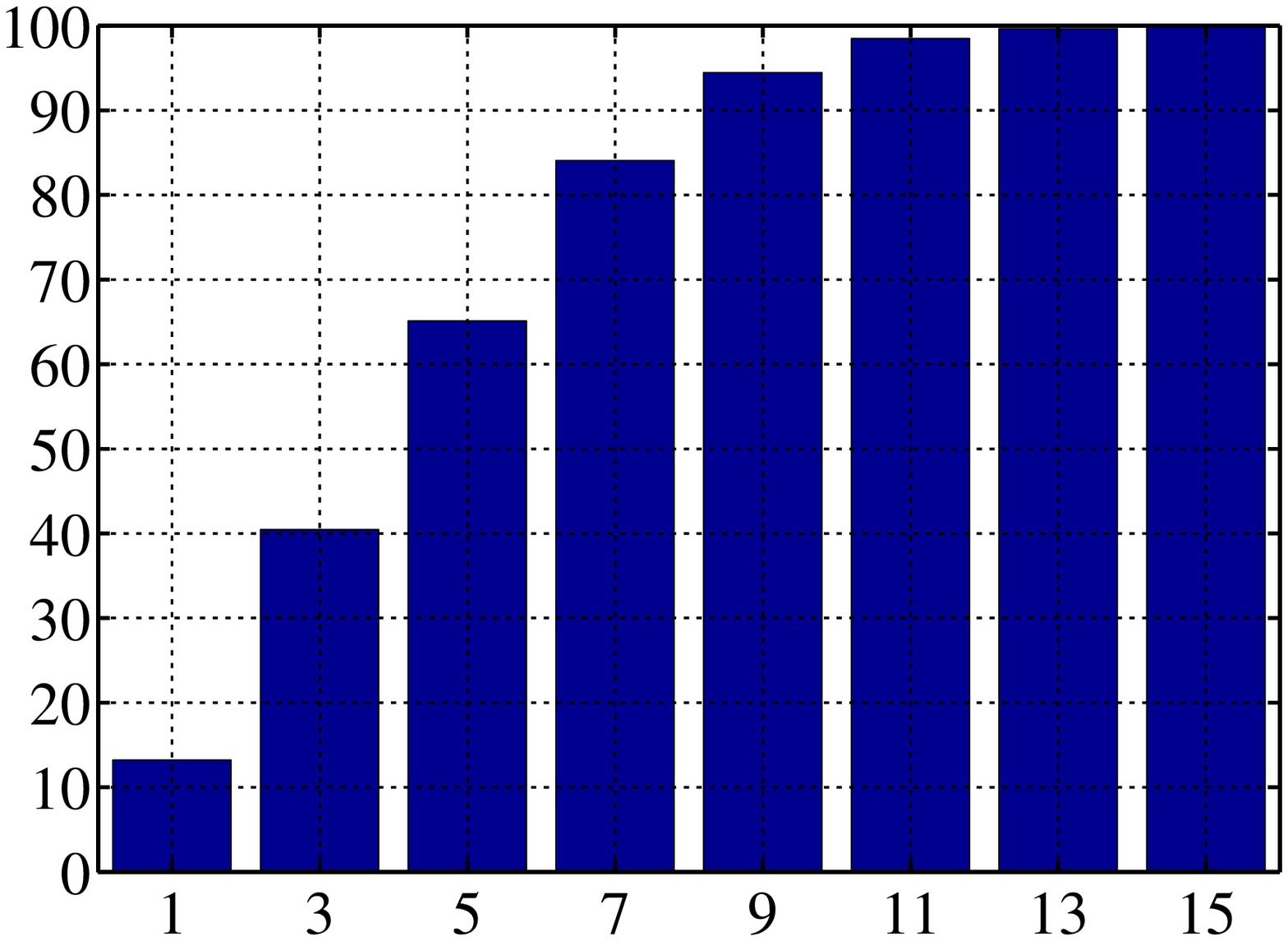} 
\put(-170,40){{\begin{sideways}{$E_{B_{2,(7,8,9)}\rightarrow C_{1,j} } \%$}  \end{sideways}}}
\put(-80,0){$j$}
% \put(-190,75){$E_{B_{2,(7,8,9)}}$}
\put(-170,120){(b)}
\caption{Energy captured by $3$ estimation sensors $C_{2,(7,8,9)}$ originates from impulse response of different number of disturbances (a) and energy harvested by using different number of outputs  $C_{1}$  from the impulse response of $3$ actuator $B_{2,(7,8,9)}$(b). The data is normalised by the maximum value when  $j=18$. The number of disturbances or outputs (elements) denotes as $j$. $j=1$ corresponds to an element located at $z=-13.5$ ($i$=7). $j=3$ corresponds to 3 elements  $i \in (6,7,8)$. The numbering convention continues the same with the central element located at $i=7$; for instance, $j=5$ corresponds to 5 elements $i \in (5,6,7,8,9)$ and so on.
 }
\label{fig:lateral}
\end{figure}
%------------------------------  end figure perturbation energy  -------------

\begin{figure}%[h!]
\begin{center}
\includegraphics[width=0.52\textwidth]{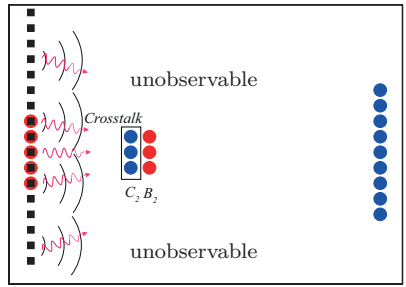}  
\put(-110,95){\small unobservable}
\put(-110,30){\small unobservable}
\end{center}
\caption{A schematic layout of the control unit. Two kind of perturbations, observed by 3 sensors $C_{2,(7,8,9)}$ are depicted; they include the perturbation coming from sources directly in front of the sensors and the lateral perturbations coming from sides which is referred to as \emph{crosstalk}.}
\label{fig:sketch_cross}
\end{figure}
%%%%%%%%%%%%%%%%%%%%%%%%%%%%%%%%%%%%%%%%%%%%%%%%%%%%%%%%%%%%%%%%%%%%%%%%%%%%%%%%%%%%%%%%%%%%%%
\subsubsection{Effect of Crosstalk}\label{crosstalk}
%%%%%%%%%%%%%%%%%%%%%%%%%%%%%%%%%%%%%%%%%%%%%%%%%%%%%%%%%%%%%%%%%%%%%%%%%%%%%%%%%%%%%%%%%%%%%%%
As a localised disturbance propagates downstream, it will -- after a short transient  -- develop into a wavepacket that grows in size and spreads along the spanwise direction. Each estimation sensor $C_{2,j}$ does not only receive a signal from the disturbance source directly upstream of it ($w_{j}$), but also the lateral sources ($w_{l}$, for $l\neq j$)  contribute to the total measured signal. The additional perturbations, originated from the lateral sources and detected by the estimation sensors $C_2$, are referred to as \emph{crosstalk} (see Fig.~\ref{fig:sketch_cross}). 

Consider now the first control unit from the previous section (3 estimation sensors and 3 actuators).  The energy of the signals received by $3$ estimation sensors from different numbers of disturbance sources $B_{1}$ is shown in Fig.~\ref{fig:lateral}a. As one can observe, around $70\%$ of the total energy of the signals originate from $3$ disturbance sources directly upstream of the estimation sensors. In order to capture $90\%$ of the total energy of the signals, $5$ disturbance sources are required in which, the additional $20\%$ of the energy belongs to the two lateral disturbance sources.  

To investigate the effect of the crosstalk in the performance of the control unit, we compare two cases. The only difference between them is the number of disturbance sources $B_1$ considered in the design process. Just as before we consider $5$ disturbance sources $B_1$ in case with crosstalk (case $D$) while we reduce the effect of crosstalk and only design the controller for $3$ disturbance sources $B_1$ ( $18/3-1-3-9/1$ or case $F$ in Tab. \ref{tab:cont_unit}). Tab. \ref{tab:cont_unit} shows the performance of the two systems; the configuration that takes into account $90\%$ of crosstalk can attenuate the energy of the disturbances up to $10.9\%$ while the configuration taking into account only  $70\%$ of crosstalk can reduce the energy up to $8.7\%$. This indicates the number of disturbance sources in the control design process depends on the nature of the disturbance (e.g. how fast it spreads in the spanwise direction). Capturing only part of the spreading of a disturbance has a sizable effect on the control performance.

Next, we investigate the performance of the controllers when the control units are replicated  along the spanwise direction. First,  we consider $6$ control units based on the configurations with high level crosstalk and with reduced-level of crosstalk. Tab. \ref{tab:cont_unit_tot} reports the reduction in the energy of the system using these controllers. The performance of $6$ control units considering the crosstalk effect (case $H$) is only $11\%$ less than the centralised controller (case $C$ in Tab. \ref{tab:pf}) where all the interconnections between the sensors and the actuators are taken into account. On the other hand, if we only capture part of the crosstalk effect (case $J$) we loose an additional $9\%$ of  performance.

%%%%%%%%%%%%%%%%%%%%%%%%%%%%%%%%%%%%%%%%%%%%%%%%%%%%%%%%%%%%%%%%%%%%%%%%%%%%%%%%%%%%%%%%%%%%%%
\subsubsection{Capturing the spread of the disturbances}
%%%%%%%%%%%%%%%%%%%%%%%%%%%%%%%%%%%%%%%%%%%%%%%%%%%%%%%%%%%%%%%%%%%%%%%%%%%%%%%%%%%%%%%%%%%%%%%
%The outputs $C_1$ are employed as an objective function to assess the performance of the controller. In fact, the ultimate objective is to suppress the amplitude of the wavepacket at the end of the domain. 

Since the wavepackets spread along the spanwise direction while propagating downstream, we need to distribute a minimum number of objective sensors $C_{1,j}$ along the span to correctly capture the energy of the disturbances. On the other hand, we have to be able to control the disturbances detected by outputs $C_1$ using the actuators $B_2$. In fact, the further away the outputs are from the centreline of an actuator, the less we can control the structures detected by that outputs. 
More specifically, we consider again control units which have 3 actuators  ($B_{2,(7,8,9)}$). Fig.~\ref{fig:lateral}b reports the energy of the signals captured by different number of outputs $C_{1}$, which originate from the impulse responses of the 3 actuators. We can observe that over $90\%$ of the total energy that originated from an impulse in the $3$ actuators is captured by 9 outputs. According to this observation, we compare two controllers, whose differentiate only in the number of employed outputs in the control design. In the first configuration (case $D$ in Tab. \ref{tab:cont_unit}) we consider $9$ outputs ($C_{1,i},\hspace{0.1cm} i=4,\cdots,12$) while in the second configuration (case $G$ in Tab. \ref{tab:cont_unit}) we implement $3$ outputs only ($C_{1,i},\hspace{0.1cm} i=7,8,9$). As one can observe in Tab. \ref{tab:cont_unit}, the reduction in the energy of the system $\bar{E_r}$ in the case with 9 outputs is $10.9\%$  while in the case with 3 outputs is $8.3\%$. 

It is important to note that in both configurations, we take into account the crosstalk effect. If we compare the performance of the controller with 3 outputs (case $G$) to the controller that only partially accounts for the crosstalk from the previous study in sec~\ref{crosstalk} (case $F$), we can observe that the energy reduction in the second case is larger, $8.3\%$ vs $8.7\%$. Finally, we compare on the performance of the $6$ control units with $9$ and $3$ outputs in  Tab.~\ref{tab:cont_unit_tot} (cases $H$ and $K$). In the former, the energy is attenuated up to $69\%$ while is the latter, it is reduced up to $48\%$.

\begin{table}
\begin{tabular}{c c c  c}
& & & \\ % put some space after the caption
\hline
\hline
Case  & Description & $\text{Norm}$     & $\text{Energy}$\\
      &             & $\text{Reduction}$ & $\text{Reduction}$\\
 $ k $& $ - $       & $1-\frac{\parallel G_{k} \parallel_2^2}{\parallel G_{n} \parallel_2^2}$ & $\bar{E_r}$\\

\hline
H   & $18/5-3-3-9/6$  & $88.0\%$&  $0.69 $\\[3pt]
J   & $18/3-3-3-9/6$  & $85.5\%$&  $0.60 $\\
K   & $18/5-3-3-3/6$  & $64.7\%$&  $0.48 $\\
\hline
\hline
\end{tabular}
\centering
\caption{In each case $6$ control unit are used. The control units distributed equidistantly along the span and does not have any overlap. The noise autocovariance for all the cases are assumed as $\alpha^2 = 10^{-6}$. In addition, the norms are computed for time $t > 2000$.}
\label{tab:cont_unit_tot}
\end{table}

%%%%%%%%%%%%%%%%%%%%%%%%%%%%%%%%%%%%%%%%%%%%%%%%%%%%%%%%%%%%%%%%%%%%%%%%%%%%%%%%%%%%%%%%%%%%%%
\section{Conclusion}
%%%%%%%%%%%%%%%%%%%%%%%%%%%%%%%%%%%%%%%%%%%%%%%%%%%%%%%%%%%%%%%%%%%%%%%%%%%%%%%%%%%%%%%%%%%%%%%
We have investigated and compared two different control strategies, namely a centralised and a decentralised. In the former approach where all the sensors and actuators are connected together, the complexity of the system (due to the number of interconnections) may be  to high for implementation in experiments, in particular, as we aim to control over a wider span of the domain. We have presented an alternative decentralised strategy, where several small control units consisting of 3 pairs of actuators-sensors are assembled to cover the full spanwise length of the flat plate. The choice 3 actuators-sensors as well as the number of source disturbances and objective sensors included in the  design of a single control unit needs to be chosen with a physical insight on the spatial and temporal scales of the perturbation inside the boundary layer. We have focused on TS wavepackets, streaky structures observed under different conditions inside the boundary layer, may need control units of different order. 

Our results reveal that the best performance is obtained for a control unit which (i) is sufficiently ``wide'' to account for the full spanwise scale of the wavepacket when it reaches the actuators and (ii) is designed to account for the perturbations which are coming from the lateral sides (crosstalk) of the estimation sensors. We may also conclude that  the influence of  crosstalk is not as essential as the spreading effect.

\section*{Acknowledgements}
The authors wish to thank Ardeshir Hanifi and Onofrio Semeraro for fruitful discussions. Computer time was provided by the Center for Parallel Computers (PDC) at the Royal Institute of Technology (KTH) and the National Supercomputer Center (NSC) at Link\"oping University in Sweden. 

% \bibliographystyle{spmpsci}      % mathematics and physical sciences
\bibliographystyle{spphys}       % APS-like style for physics
%\bibliography{./refs,gaster}   % name your BibTeX data base
\bibliography{refs,gaster}   % name your BibTeX data base
\end{document}